\documentclass[fleqn,usenatbib]{mnras}

\usepackage{newtxtext,newtxmath}

\usepackage[T1]{fontenc}
\usepackage[utf8]{inputenc}

\DeclareRobustCommand{\VAN}[3]{#2}
\let\VANthebibliography\thebibliography
\def\thebibliography{\DeclareRobustCommand{\VAN}[3]{##3}\VANthebibliography}

\usepackage{graphicx}	
\usepackage{amsmath}	

\title[Variable stars in M71]{Variable stars in the field of the Galactic globular cluster M71}

\author[C. C. Cort\'es et al.]{
C. C. Cort\'es$^{1}$\thanks{E-mail: caddy.cortes@ucentral.cl},
D. Deras $^{2,3},$
A. Arellano Ferro $^{4}$\thanks{Corresponding author: armando@astro.unam.mx},
S. Muneer $^{5}$,
and I. H. Bustos Fierro $^{6}$
\\
$^{1}$Centro de Investigaci\'on en Ciencias del Espacio y F\'isica Te\'orica, Universidad Central de Chile, Av. Francisco de Aguirre 0405, La Serena, Chile\\
$^{2}$ Instituto Nacional de Astrof\'isica, \'Optica y Electr\'onica (INAOE), Luis Enrique Erro No.1, C.P. 72840, Tonantzintla, Pue., M\'exico\\
$^{3}$ Secretaría de Ciencia, Humanidades, Tecnolog\'ia e Innovaci\'on (Secihti), Av. de los Insurgentes Sur 1582, C.P. 03940, Ciudad de México, M\'exico\\
$^{4}$ Instituto de Astronom\'ia, Universidad Nacional Autonoma de M\'exico, Ciudad Universitaria, C.P. 04510, Ciudad de M\'exico, M\'exico\\
$^{5}$ Indian Institute of Astrophysics, 2nd block Koramangala, Bangalore, 560 034, India\\
$^{6}$ Observatorio Astron\'omico, Universidad Nacional de C\'ordoba, X5000IND C\'ordoba, Argentina
}

\date{Accepted XXX. Received YYY; in original form ZZZ}

\pubyear{\the\year{}}

\begin{document}
\label{firstpage}
\pagerange{\pageref{firstpage}--\pageref{lastpage}}
\maketitle

\begin{abstract}
M71 is a nearby, metal-rich globular cluster at low Galactic latitude, where field contamination and spatially variable extinction complicate colour–magnitude diagrams (CMDs) and the identification of cluster member variable stars. Our aims are (i) to construct a homogeneous census of variable stars in M71 by refining their periods and classifications and identifying new candidates, and (ii) to derive a decontaminated, differentially dereddened CMD to constrain its physical properties.
We obtained Johnson–Kron–Cousins $VI$ time-series CCD photometry and reduced it using difference image analysis. Cluster membership was established from \textit{Gaia}~DR3 proper motions, and a differential-reddening correction was applied across the field of view. The resulting CMD, cleaned of field stars, was compared with tailored isochrones to estimate age ($12.9^{+0.9}_{-0.8}$ Gyr), metallicity ([Fe/H] =$-0.88^{+0.13}_{-0.15}$), mean reddening ($E(B-V)$ = $0.21 \pm 0.02$), and distance modulus ($(m-M)_{0}$ = $13.01 \pm 0.06$). Variable stars were identified using two complementary approaches: a periodogram-free string-length scan refined with phase dispersion minimisation, and a robust inter-site screening based on median statistics combined with a generalised Lomb–Scargle significance criterion.
We identified 21 variable stars not previously reported in the Catalog of Variable Stars in Globular Clusters and provided their periods, amplitudes, classifications, membership status, and light curves. This combined strategy yields a consistent picture of M71, expanding its known variable-star population and confirming parameters typical of metal-rich Galactic disk globular clusters.
\end{abstract}




\section{Introduction}

Globular clusters (GCs) are fundamental laboratories for investigating the formation history and chemical evolution of the Milky Way. Among them, M71 (NGC 6838) is a low-density and sparse GC, and it stands out due to its relatively high metallicity, proximity, and low Galactic latitude, located at $l = 56^{\circ}.75$, $b = -4^{\circ}.56$. With a mean metallicity of $[{\rm Fe/H}] = -0.82 \pm 0.02$ dex \citep{Carretta2009}, M71 is classified as a metal-rich GC and is frequently used as a benchmark for studying chemical enrichment processes in the Galactic disk \citep{Alves-Brito2008,DiCecco2015}. Also, it is believed to have been formed \textit{in situ}, belonging to the Galactic disk \citep{Callingham2022}. 

Its heliocentric distance \citep[$\sim$4 kpc,][]{Grundahl2002,Baumgardt2021} and moderate reddening \citep[$E(B-V) \sim 0.25$,][]{Harris1996} make it accessible for detailed photometric and spectroscopic studies. 
This GC is relatively young, with an estimated age between 9 and 11 Gyr, although \citet{DiCecco2015} report an absolute age of $12 \pm 1$ Gyr using deep optical photometry.

It is likely that this GC is following a disk-type orbit \citep{Geffert2000} which suggests that it has lost a substantial fraction of its initial mass due to gravitational interactions with the Galactic field, as well as dynamical collisions caused by encounters with molecular clouds and/or spiral arms \citep{Cadelano2017}. Currently, its total mass is estimated to be 2 $\times$ 10$^{4}M_{\odot}$ \citep{Kimmig2015}. 

The Catalogue of Variable Stars in Globular Clusters (CVSGC; \citealt{Clement2001})\footnote{https://www.astro.utoronto.ca/~cclement/cat/C1951p186} lists 29 variables, composed primarily of binaries, SX Phoenicis (SX Phe) stars, and semi-regular (SR) variables, and a 5 ms pulsar whose membership status is unknown. At the relatively high metallicity of M71, the absence of confirmed member RR Lyrae stars is not unusual among globular clusters with predominantly red Horizontal Branch (HB) morphologies. In this sense, M71 is consistent with other relatively metal-rich clusters, whose colour–magnitude diagrams (CMDs) are typically characterised by a poorly populated instability strip and a well-developed red clump.

\citet{SawyerHogg1973} reported V4 as a potential RRc star in the field of the cluster but it was later dismissed as a field star \citep{Curdworth1985,PrudilArellanoFerro2024}. One of the aims of this paper is to update the census of variable stars in M71 in the light of recent investigations and our present data.

The paper is organised as follows. Sect.~\ref{sec:data} describes the observations, Difference Image Analysis (DIA) processing, and photometric calibration; Sect.~\ref{sec:membership} details the membership selection from \textit{Gaia}~DR3 and the construction of the decontaminated CMD; Sect.~\ref{sec:diff_red} presents the differential-reddening correction to the CMD; Sect.~\ref{sec:cmd} discusses the CMD and the isochrone-based inference of age, metallicity, distance, and reddening; Sect.~\ref{sec:var_stars} presents the two-pronged variability search, the cross-match with the Clement catalogue, and the refined periods; Sect.~\ref{sec:conclusions} summarises our findings and perspectives.

\section{Data and Reduction} \label{sec:data}

\subsection{Observations} \label{subsec:obs}
We obtained Johnson–Kron–Cousins $VI$ imaging at two sites. 
The first dataset was acquired at the San Pedro Mártir Observatory in Baja California, Mexico, using the 0.84\,m telescope, from 2017 October 30 to November 6.
The telescope was equipped with a Spectral Instruments 2048 $\times$ 2048 CCD at $0\farcs444$\,px$^{-1}$, yielding an effective field of view of approximately $7\farcm4\times7\farcm4$ per frame for our monitoring setup.

The second dataset was acquired at the Indian Astronomical Observatory (IAO; Hanle, India) on 2018 October 14–15 and 2019 October 24–25, using the 2\,m telescope equipped with an E2V CCD44-82-0-E93 detector ($2048 \times 4096$\,px; $0\farcs296$\,px$^{-1}$). 
Time-series photometry used a $310\times310$\,px subarray, while the stacked reference image employed for the finding chart was built from full-frame exposures, delivering a field of view of approximately $10\farcm1\times10\farcm1$. Table \ref{log} presents the observing log, including the number of images, typical exposure times, and the average seeing conditions.

\begin{table}
\footnotesize
\caption {Log of the observation of M71 (NGC 6838)$^*$.}
\label{log}
\centering
\begin{tabular}{ccccccc}
\hline
Date & Site	& $N_V$ &$t_V (s)$ &$N_I$ & $t_I (s)$ &Avg.\\
   &   &          & sec    &         &sec&seeing (")\\
\hline
2017-10-29	& SPM	& -- & -- & 18 & 10 & 2.20 \\
2017-10-31	& SPM & 1 & 120 & 1  & 50 & 1.85  \\
2017-11-02	& SPM & -- & -- & 2  & 80 & 1.75  \\
2017-11-03	& SPM & 5 & 120 & 8  & 80 & 2.16  \\
2017-11-05  & SPM & -- & -- & 1  & 80 & 2.25  \\
2017-11-06	& SPM & 4 & 120 & 10  & 80 & 1.96 \\
2018-10-14	& HANLE	& 33 & 15 & 31 & 5 & 2.14 \\
2018-10-15	& HANLE & 74 & 15 & 76  & 5 & 2.30  \\
2019-10-24	& HANLE & 24 & 10 & 22  & 2.5 & 1.52  \\
2019-10-25	& HANLE & 32 & 10 & 28  & 2.5 & 1.75  \\

\hline
Total:&  &  173 &  & 197  &  &   \\
\hline
\end{tabular}
\raggedright
\quad $*$
Columns $N_V$ and $N_I$ correspond to the number of images obtained. $t_V$ and $t_I$ indicate the typical exposure times. The average nightly seeing is given in the last column.
\end{table}

\subsection{Difference Image Analysis} \label{subsec:image}

The images were calibrated using standard bias subtraction and flat-fielding. To obtain high-precision time-series photometry in the crowded field of M71, we adopted DIA.

Data reduction was performed with the \textsc{DanDIA}\footnote{\textsc{DanDIA} is built from the DanIDL library of IDL routines: \url{http://www.danidl.co.uk}.} pipeline \citep{Bramich2013}, which implements a discrete pixel array to model the convolution kernel used to match the point-spread function (PSF) between images of the same field \citep{Bramich2008}. A comprehensive description of the method is provided by \citet{Bramich2011}.

Briefly, a high signal-to-noise reference image is constructed by stacking several best-seeing exposures. Each science image is then convolved with a spatially varying kernel to match the PSF and subtracted from the reference. The differential flux of each star is measured on the resulting differential image, and the total flux at epoch $t$ is recovered as
\begin{equation}
f_{\mathrm{tot}}(t) \;=\; f_{\mathrm{ref}} \;+\; \frac{f_{\mathrm{diff}}(t)}{p(t)} ,
\label{eq:flux}
\end{equation}
where $f_{\mathrm{ref}}$ is the reference flux (ADU\,s$^{-1}$), $f_{\mathrm{diff}}(t)$ is the differential flux measured on the subtracted image at epoch $t$, and $p(t)$ is the photometric scale factor (i.e. the integral of the kernel solution over the PSF).

The instrumental magnitude at each epoch is then
\begin{equation}
m_{\mathrm{ins}}(t) \;=\; 25.0 \;-\; 2.5 \,\log_{10}\!\big[f_{\mathrm{tot}}(t)\big] ,
\label{eq:mag}
\end{equation}
where the additive constant sets the instrumental zero point. This DIA workflow minimises systematics due to PSF variations and blending, enabling millimagnitude-level precision in crowded fields.

\subsection{Transformation to the standard system}

Local standard stars in the field of M71 from the online collection of \citet{Stetson2000}\footnote{\url{http://www3.cadc-ccda.hia-iha.nrc-cnrc.gc.ca/community/STETSON/standards}} were used to transform our instrumental $vi$ magnitudes onto the Johnson--Kron--Cousins $VI$ system. A total of 208 and 143 local  standards were employed to define the transformation relations for Hanle and SPM, respectively. 

\begin{figure*}
  \centering
  \includegraphics[width=\hsize]{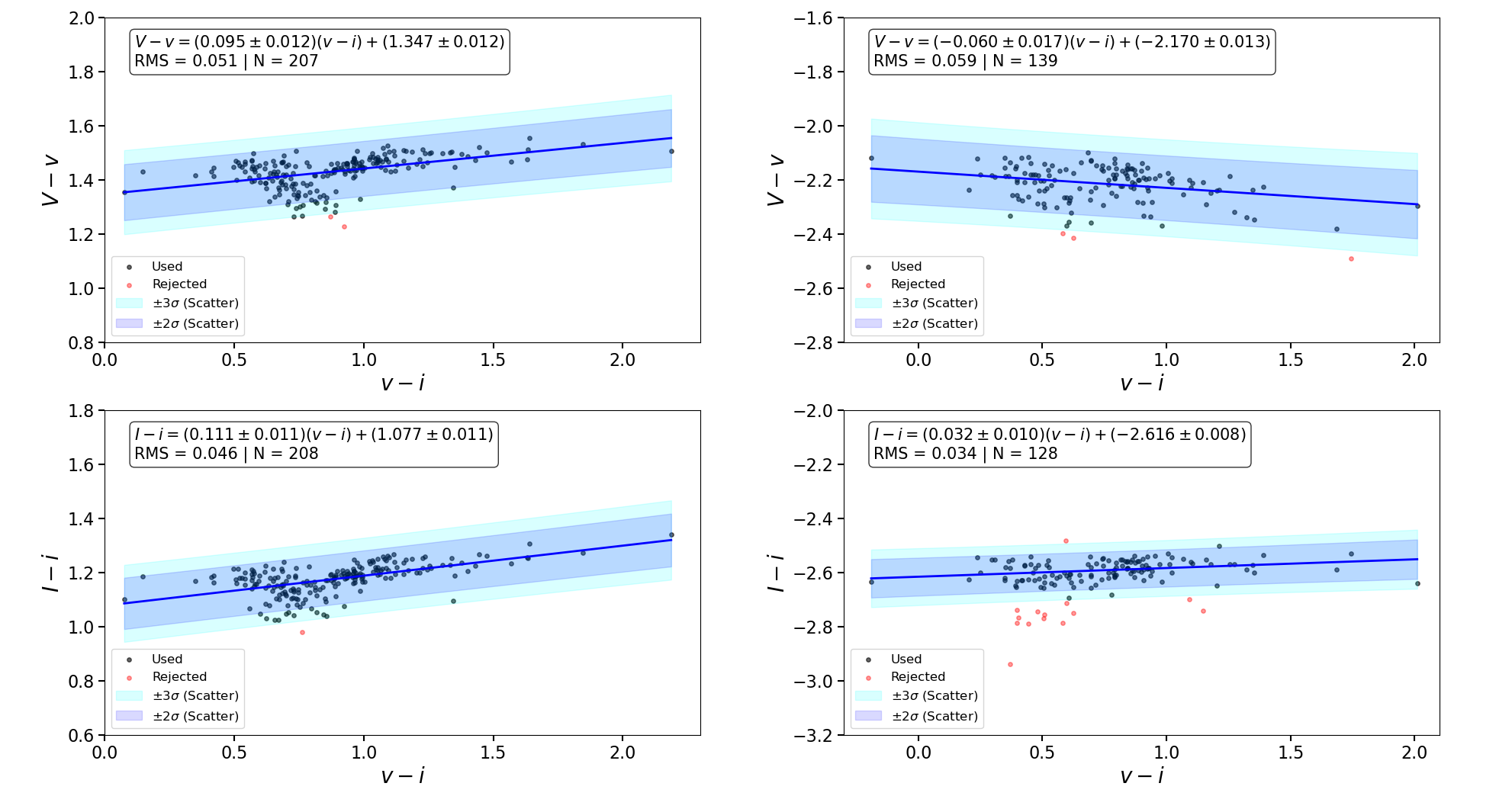}
\caption{Transformation equations from instrumental to standard $VI$ magnitudes using 208 and 143 local standards from \citet{Stetson2000} for Hanle and SPM, respectively. \textit{Left panel}: Hanle; \textit{Right panel}: SPM. The panels list the fitted coefficients (with 1$\sigma$ uncertainties) and the rms of the residuals. To further illustrate the reliability of the fits, we include shaded regions representing the $\pm$2$\sigma$ (dark blue) and $\pm$3$\sigma$ (light blue) confidence intervals around the best-fit relations.}
  \label{Trans_color}
\end{figure*}

\begin{figure}
  \centering
  \includegraphics[width=\hsize]{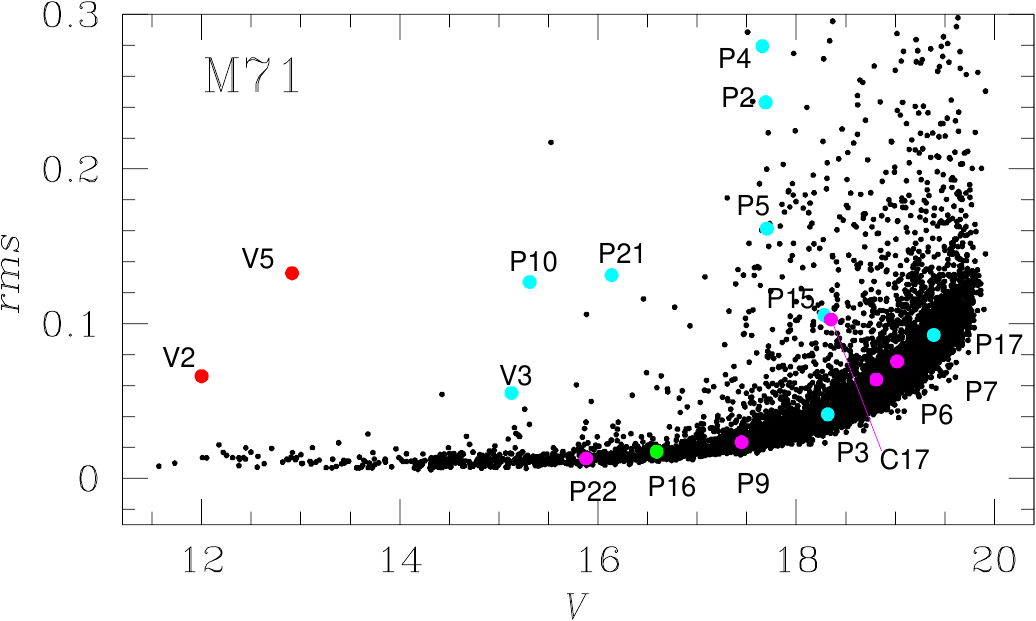}
\caption{The rms $V$ magnitude deviations as a function of mean $V$ magnitude. Variables listed in the CVSGC measured by our photometry are labeled colour coded as: red-SR, green-RRc, cyan-eclipsing binaries and magenta-SX Phe. See Sect.~\ref{sec:var_stars} for comments on the variability detection and type assignation.}
  \label{fig:rms}
\end{figure}

We adopted linear colour equations derived by least-squares fitting, including colour terms only when statistically significant. Using lower-case letters for instrumental magnitudes and upper-case for standard magnitudes, the relations take the form:
\begin{align}
V &= v + \alpha_V + \beta_V\,(v-i), \\
I &= i + \alpha_I + \beta_I\,(v-i),
\end{align}
where $\alpha_{V,I}$ are zero points and $\beta_{V,I}$ are colour coefficients. We performed an iterative 2$\sigma$ and 3$\sigma$ clipping procedure based on the residuals of the linear fits to ensure that the transformation equations are not affected by outliers. In this process, stars deviating by more than 3$\sigma$ are removed, and the fit is recomputed until convergence is achieved. The coefficients (with $1\sigma$ uncertainties) and the rms of the residuals are reported in the panels of Fig.~\ref{Trans_color}. These transformations were then applied to all instrumental light curves. In Fig.~\ref{fig:rms} we show the photometric quality of our observations. At bright magnitudes ($V \leq 14$–15), the rms scatter reaches a well-defined floor of $\lesssim$ 0.01–0.02 mag, indicative of high signal-to-noise measurements. Towards fainter magnitudes, the rms increases gradually, following the expected behavior driven by photon noise. By $V \approx 18$–20, the scatter rises to $\gtrsim$ 0.05–0.1 mag, marking the transition to the noise-limited regime at low flux levels.

\begin{figure*}
  \centering
  \includegraphics[width=\hsize]{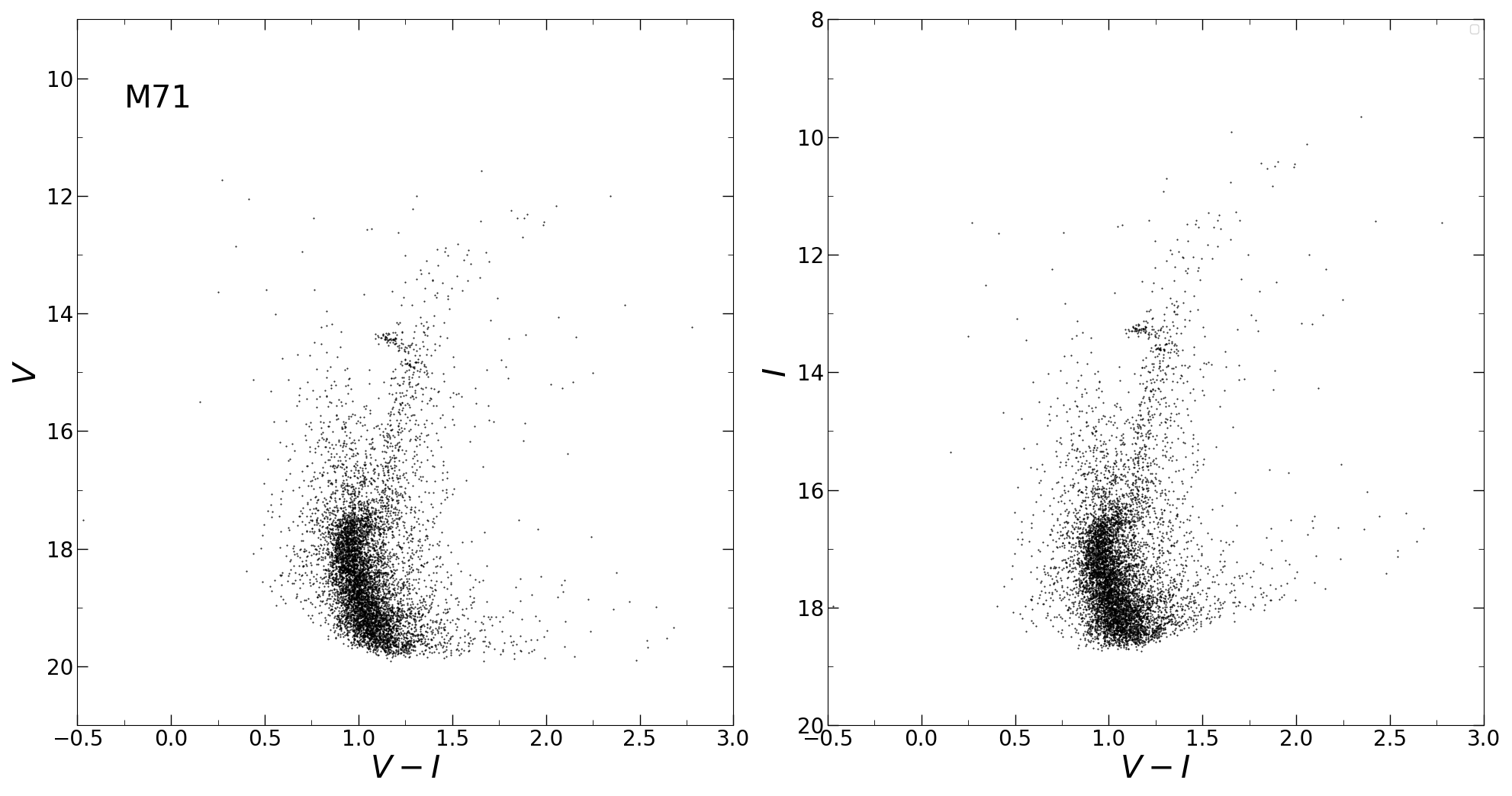}
  \caption{CMD of M71 from the Hanle dataset used in this work. Left: $V$ versus $(V-I)$. Right: $I$ versus $(V-I)$. Both panels use the same photometric sample after transformation to the standard system.}
  \label{fig:M71_CMD}
\end{figure*}


\section{Stellar membership analysis}
\label{sec:membership}

Because M71 lies close to the Galactic bulge, its CMD is significantly affected by field interlopers (see Fig. \ref{fig:M71_CMD}). To mitigate this, we performed a decontamination based on \textit{Gaia}~DR3 proper motions \citep{Gaia2023}, following the two–stage procedure of \citet{Bustos2019}. 

In the first stage, we applied the BIRCH algorithm \citep{Zhang1996} in a four–dimensional space defined by the gnomonic projection of the sky coordinates and the two proper–motion components. This step clusters stars in the 4D parameter space. In the second stage, we examined the projected spatial distribution of stars as a function of proper motion, which allows us to flag likely cluster members—including stars in the outskirts and those with relatively large proper–motion dispersion.

\begin{figure*}
\begin{center}
\includegraphics[width=16cm]{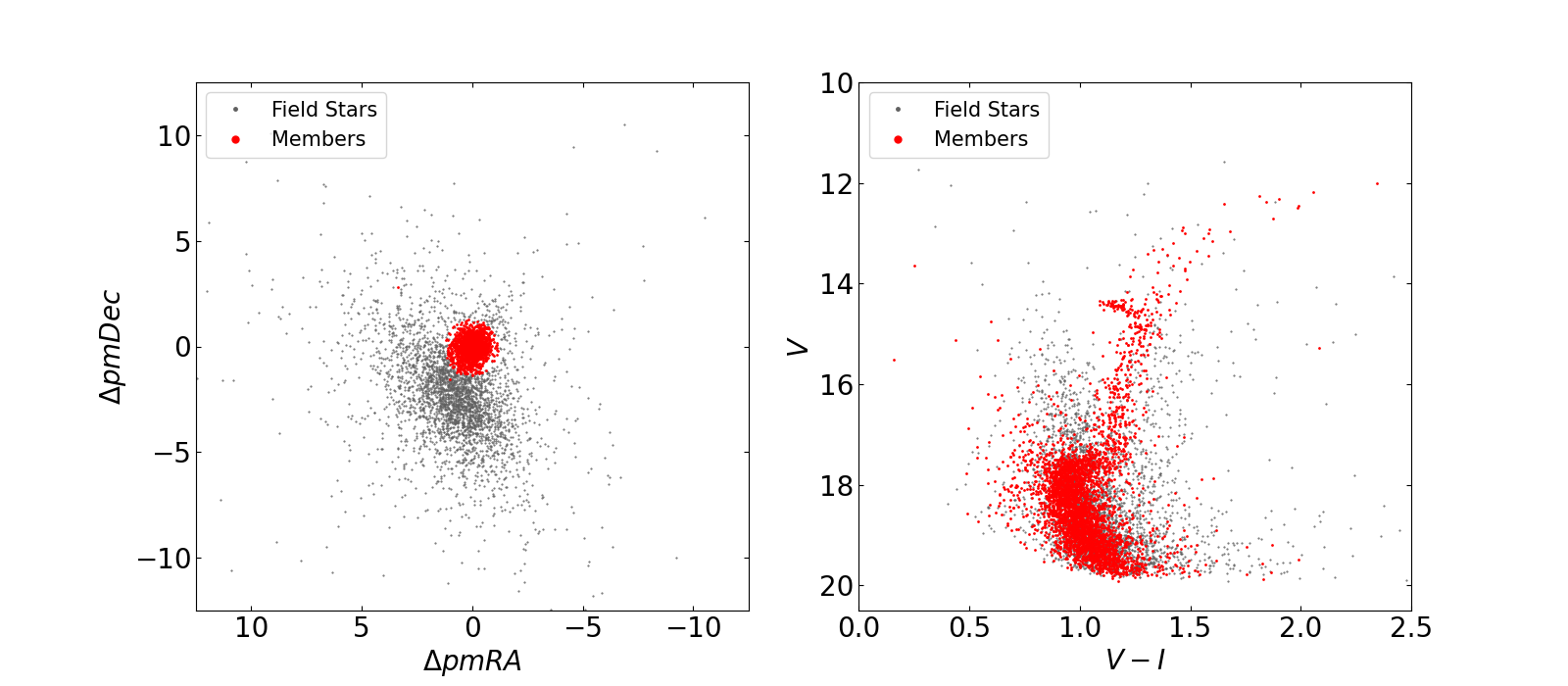}
\caption{Red and gray symbols mark likely cluster members and field stars, respectively, as determined in Sect.~\ref{sec:membership} from a cross–match between the \textit{Gaia}~DR3 catalogue and our photometry. \textit{Left panel:} Vector–point diagram (VPD) of the likely members in proper–motion space, referred to the cluster motion $\mu_{\alpha}\cos\delta=-3.412\pm0.011$ and $\mu_{\delta}=-2.662\pm0.011$ \citep{Baumgardt2021}. \textit{Right panel:} Decontaminated CMD of M71. Within our FoV we find 6\,733 \textit{Gaia} sources with measured proper motions, of which 3\,743 are likely cluster members.}
\label{fig:Members}
\end{center}
\end{figure*}

Candidate members were then validated by requiring consistency across three diagnostics: (i) their sky positions, (ii) their locus in the vector–point diagram (VPD) of proper motions, and (iii) their placement in the CMD, all compatible with a globular cluster sequence. Finally, we cross–matched the validated members with our photometric catalogue to assign the corresponding \textit{Gaia}~DR3 proper motions (see Fig.~\ref{fig:Members}). 

After decontamination, the main CMD features become clearly discernible. The main sequence spans $19.9 \lesssim V \lesssim 17.8$ with a turn–off at $V=17.6$. The CMD also shows a horizontal branch at $V=14.5$, predominantly on the red side (red clump), a well–defined subgiant and red–giant branches, and the characteristic red bump at $V=14.9$.

The membership analysis presented here is based on \textit{Gaia}~DR3 proper motions and therefore inherits the known limitations of Gaia in the most crowded central regions of globular clusters. In particular, source detection and astrometric quality may be reduced towards the cluster centre, implying that some stars identified through our DIA photometry may lack reliable \textit{Gaia} counterparts or may not be included in the \textit{Gaia}-based membership selection. Consequently, the membership analysis should not be interpreted as a completeness assessment of the variable-star census in the central region, but rather as the best membership characterisation currently available for the subset of stars with usable \textit{Gaia} astrometry.

\section{Differential reddening correction}
\label{sec:diff_red}

A major challenge in the study of this GC is the accurate determination of its interstellar reddening. Its reported value is $E(B - V)$ = 0.25 \citep{Harris1996}, a consequence of its lower Galactic latitude and proximity to the Galactic bulge. The evolutionary sequences in the CMDs shown in Fig. \ref{fig:M71_CMD} appear noticeably distorted or broadened, suggesting that differential reddening may be present. Quantifying this effect is essential as it can introduce significant variations in the shape and positions of key characteristics of the CMD, potentially leading to incorrect estimates of distance, metal abundance, and age. Starting from the selection of likely member stars derived in the previous section, and at a fixed magnitude range ($12.0 < V < 20.0$) we divided the CMD into magnitude bins of 0.5 mag, except at the level of the MS-TO and SGB ($16.5 < V < 19.0$), where we used 0.15 mag bins to obtain a finer sampling. For each magnitude bin, we computed the 3$\sigma$-clipped median values of the ($V - I$) colour and the $V$ magnitude. These medians were interpolated to create a mean ridge line (MRL), which is used as a reference for estimating the geometric distance $\Delta X$ of each likely cluster member, projected along the direction of the reddening vector. By assuming the standard extinction coefficient $R_V = 3.1$, this vector is defined by the extinction coefficients $R_{V} = 3.103$ and $R_{I} = 1.857$, obtained from \citet{Cardelli1989}.
For each likely cluster member, we identified the $n$ nearest reference stars and computed the 3$\sigma$-clipped median of their geometric distances to the MRL, along the reddening vector. This median value corresponds to the star's assigned displacement $\Delta X$, which is transformed into the relative differential reddening $\delta E(B-V)$ by using:
\begin{equation}
\delta E(B-V) = \frac{\Delta X}{\sqrt{R_{V}^{2}+(R_{V}-R_{I})^{2}}}
\end{equation} 
To increase the spatial resolution we iteratively performed this computation three times using the $n$ = 60, 40, and 20 closest stars to each likely cluster member. The differential variations of the colour excess within the sampled field of view are significant, ranging between $-0.07 < \delta E(B-V) < 0.09$.
To determine whether the broadening of the observed CMD is primarily caused by differential reddening, we quantified the contribution from photometric uncertainties. We first selected a sample of stars with rms values within 3$\sigma$ ($\lesssim$ 0.17 mag) to minimise the impact of poorly measured sources. Using this sample, we constructed the CMD and divided it into bins of 0.5 mag in $V$. For each bin, we computed the 3$\sigma$-clipped median colour and magnitude to define representative centroids, which were then interpolated to create the ridge line of the CMD. We focused on the MSTO region (17.0 $< V <$ 18.0), where the effects of differential reddening are expected to be most pronounced. In this magnitude range, we measured a colour dispersion relative to the ridge line of $\sigma_{\rm obs} \approx 0.11$ mag, while the typical photometric uncertainty is $\sigma_{\rm phot} \approx 0.05$ mag (1$\sigma$), corresponding to $\sim$0.33 mag and $\sim$0.15 mag at the 3$\sigma$ level, respectively. After subtracting the photometric contribution, the intrinsic dispersion remains $\sigma_{\rm int} \approx 0.11$ mag. This indicates that photometric uncertainties account for $\sim$19$\%$ of the total variance, implying that the observed CMD broadening is dominated by differential reddening.

\begin{figure*}
\centering
\includegraphics[width=\linewidth]{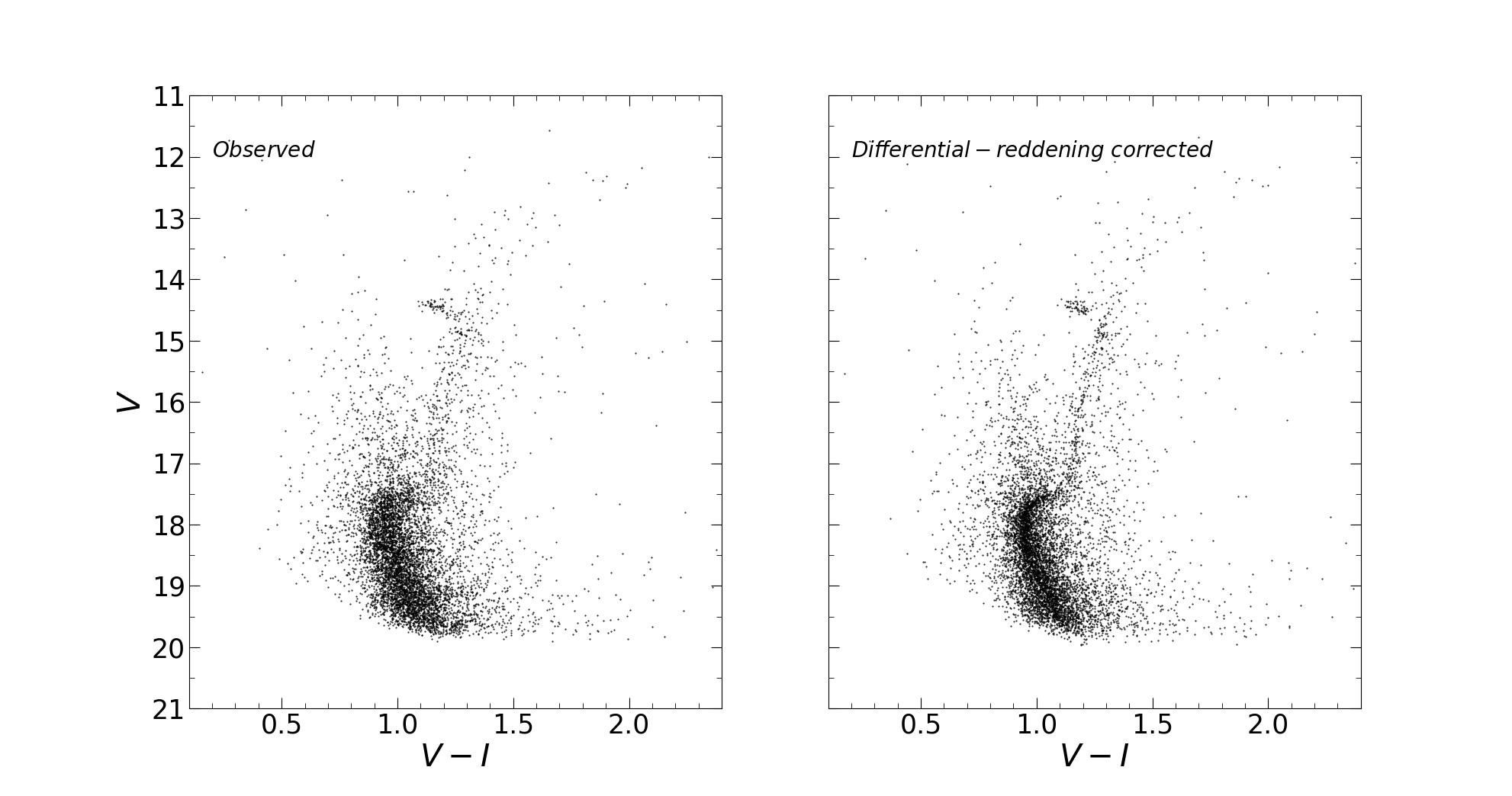}
\caption{Comparison between the original CMD from the Hanle observations used in this work (left) and the CMD after differential–reddening correction (right).}
\label{fig:diff_red}
\end{figure*}

\begin{figure*}
\centering
    \includegraphics[width=\textwidth]{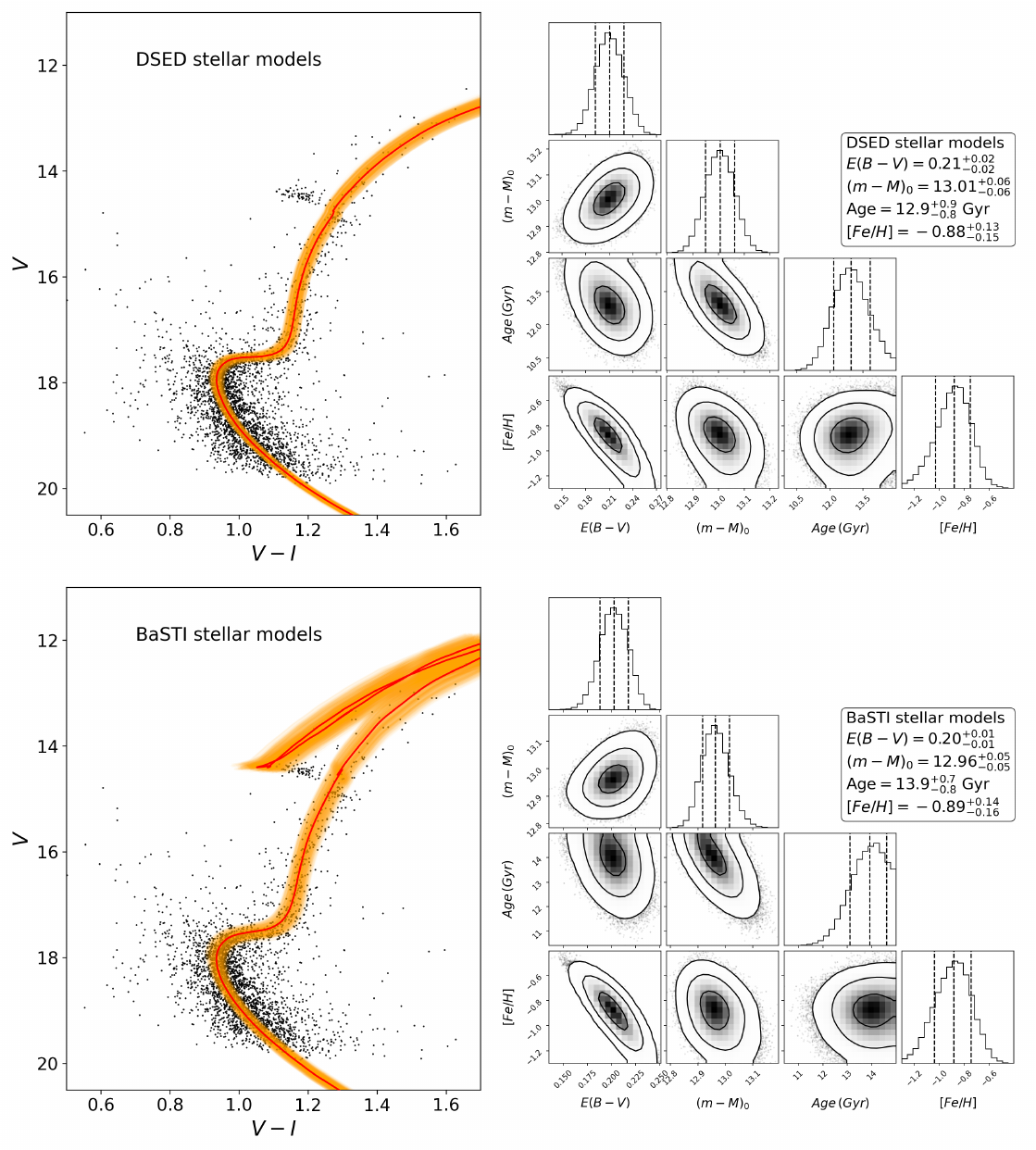}    
\caption{Isochrone fitting of the M71 CMD corrected for differential reddening. \emph{Left:} best–fit DSED (top) and BaSTI (bottom) isochrones (red) with $1\sigma$ envelopes (orange). \emph{Right:} corner plots showing the one– and two–dimensional posterior projections for all fitted parameters (contours at $1\sigma$, $2\sigma$, $3\sigma$).}
\label{fig:isochrones}
\end{figure*}






\section{The color magnitude diagram}
\label{sec:cmd}

\subsection{Isochrone fitting}

The decontamination of the CMD of M71 using stellar PMs derived by \textit{Gaia}~DR3, along with the differential reddening correction described in the previous sections, allowed us to construct a clean catalogue of likely cluster members. This refined sample is ideally suited to carry out a tightly constrained photometric estimation of the fundamental parameters of the cluster, including  absolute age, metallicity, distance modulus, and reddening. To this end, we have used a Bayesian approach akin to that used by \citet[][see also \citealp{Saracino2019,Cadelano2019,Cadelano2020b,Deras2023,Deras2024}]{Cadelano2020a} which is based on performing an isochrone fitting of the CMD derived from the cleaned star catalogue. We performed the fit only on stars in the magnitude range $15.5 < V < 19.0$ covering the TO and the SGB, and just below the RGB-bump, given that this region is the most sensitive to metallicity and stellar age variations. The isochrones were acquired from two databases, namely the Dartmouth Stellar Evolution Database (DSED) \citep{Dotter2008}\footnote{\url{http://stellar.dartmouth.edu/models/isolf_new.html}}, and the Bag of Stellar Tracks and Isochrones (BaSTI) \citep{Pietrinferni2021}\footnote{\url{http://basti-iac.oa-abruzzo.inaf.it/index.html}}. For each isochrone, we assumed a standard He content $Y$ = 0.25, and $[\alpha \rm /Fe]$ = +0.4, a typical value for bulge GCs. For each isochrone dataset, we created a grid spanning  ages from 9.0 to 15.0 Gyr, metallicities from $[\rm Fe/H]=-1.30$ to $[\rm Fe/H]=-0.40$, distance moduli between 12.5 and 13.5, and reddening values between 0.1 and 0.5. For these parameters, we assumed a Gaussian priors within their respective range of values. To find the best possible fit, we compared our CMD to the isochrone grid using a Markov Chain Monte Carlo (MCMC) sampling technique, as detailed in \citet[][see their Section 4.2]{Cadelano2020a}. To sample the posterior probability distribution in the $n$-dimensional parameter space, we used the \texttt{emcee} code \citep{Foreman2013,Foreman2019}. Temperature-dependent extinction coefficients from \citet{Casagrande2014} were used when converting absolute magnitudes to the observed frame. For the reddening, metallicity, and distance modulus, we adopted prior Gaussian distributions centered at $E(B-V)$ = $0.25 \pm 0.03$ \citep{Harris1996}, $\rm [Fe/H] = -0.82 \pm 0.05$ \citep{Carretta2009}, and DM = $13.01 \pm 0.01$ \citep{Grundahl2002,Baumgardt2021}, respectively. The left panels in Fig. ~\ref{fig:isochrones} show the observed CMD and the best-fit isochrones with their 1$\sigma$ uncertainties for each of the two adopted sets of theoretical models. At this point, we emphasize that different stellar evolution models can yield systematically different age estimates for globular clusters, with discrepancies of up to $\sim$1-2 Gyr depending on the adopted input physics and chemical assumptions (e.g., \citet{Gontcharov2020}). Despite the 1 Gyr offset in the absolute age, the best-fit isochrones properly reproduce the evolutionary sequences of our observed CMD, which are consistent within their respective uncertainties. The one- and two-dimensional posterior probabilities for all of the parameter combinations are presented on the right panels as corner plots. The best-fit values and their uncertainties (based on the $16^{th}$, $50^{th}$, $84^{th}$ percentiles) are also summarised in Table \ref{tab:iso_best}.

\begin{table*}
        \begin{center}
                \caption{Best-fit values of age, metallicity, colour
                  excess and absolute distance modulus obtained from
                  isochrone fitting.}\label{tab:iso_best}
                \begin{tabular}{|c|c|c|c|c|} 
                        \hline
                        Model   &   Age    & $[\rm Fe/H]$ & $E(B-V)$   & $(m-M)_{0}$ \\
                                & [Gyr]   &   dex &    [mag]    &  [mag]     \\                  
                        \hline
                        DSED   &  $12.9^{+0.9}_{-0.8}$ & $-0.88^{+0.13}_{-0.15}$ & $0.21^{+0.02}_{-0.02}$ & $13.01.^{+0.06}_{-0.06}$  \\ \\
                        BaSTI  &  $13.9^{+0.7}_{-0.8}$ & $-0.89^{+0.14}_{-0.16}$ & $0.20^{+0.01}_{-0.01}$ & $12.96^{+0.05}_{-0.05}$  \\ 
                        \hline
                \end{tabular}
        \end{center}
\end{table*}

\begin{figure*}
    \centering
    \includegraphics[width=1.0\textwidth]{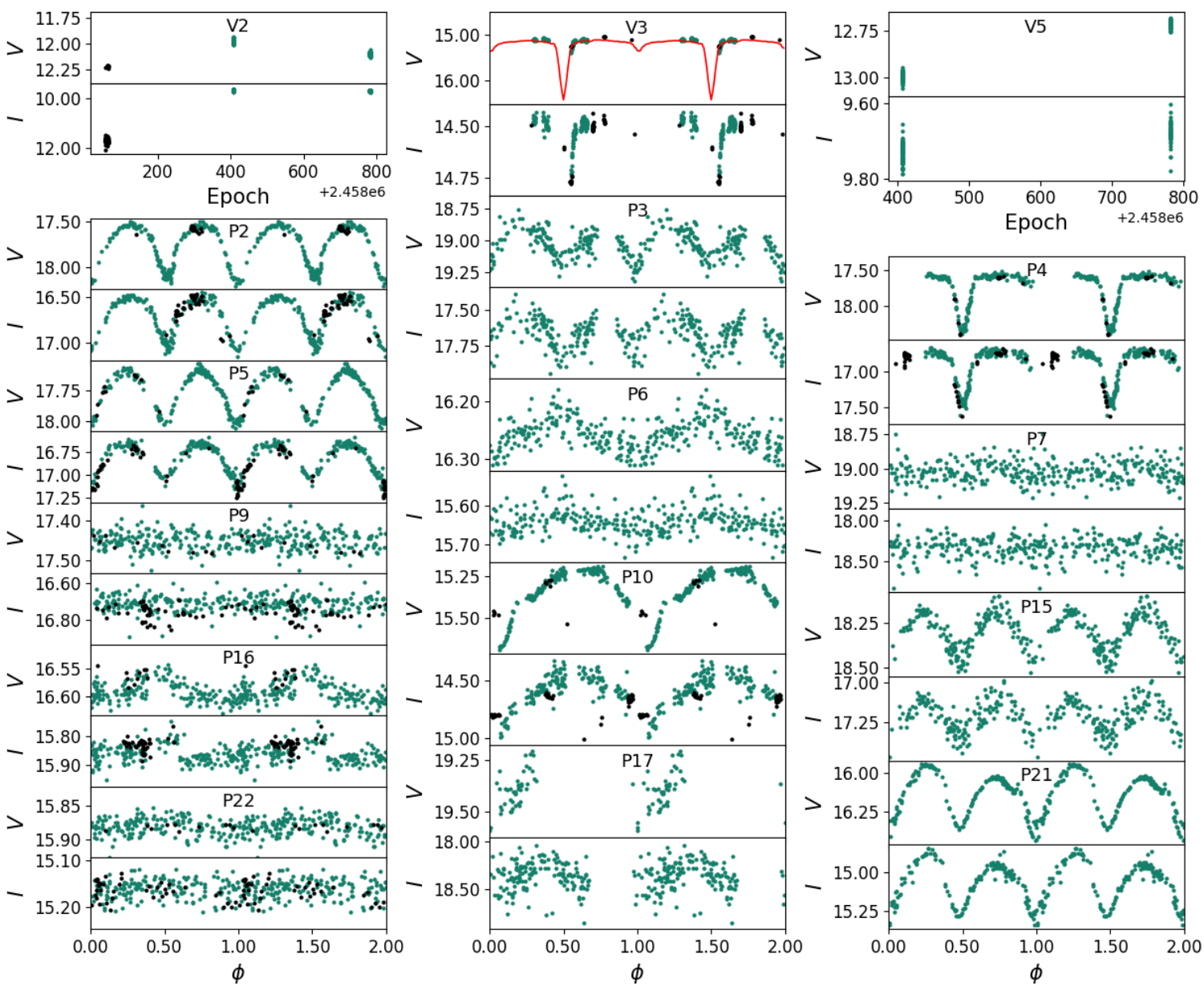}
    \caption{Light curves of the previously reported variables listed in Table~\ref{tab:clement_comparison}. For V2 and V5, the available data are shown as a function of epoch, since no reliable period could be derived from our photometry. For the remaining stars, the light curves are phased with the adopted periods $P_V$ listed in column 8 of Table~\ref{tab:clement_comparison}. In each panel, the upper and lower subpanels correspond to the $V$- and $I$-band data, respectively. Hanle measurements are shown in dark teal, while SPM measurements are plotted in black. For V3, the orbital solution from \citet{Jeon2006} is overplotted as a red line.}
    \label{fig:Clement_Variable}
\end{figure*}

\begin{table*}
\centering
\caption{Comparison of periods for variable stars in M71 reported in the CVSGC ($ P_{C}$) and updated in this work ($ P_{V}$). The numbers in parenthesis represent the uncertainty in the last decimal places.}
\label{tab:clement_comparison}
\begin{tabular}{lccccccccc}
\hline
$\rm ID_{C}$ & RA  & Dec   & $\overline{V}$  & $\overline{I}$ & Epoch min.& $ P_{C}$  & $  P_{V}$ & $A_{V}$ & $A_{I}$  \\
 &(J2000)&(J2000)& mag &mag& 2450000.+&day & day & mag & mag \\
\hline
V2   &19:53:49.41&+18:44:26.79 & 11.87 &9.53 & 8781.1682& 200.& ... & ... &...\\
V3   &19:53:49.34&+18:45:43.23 & 15.18 &14.55 &8782.0959 & 3.7908 & 0.707128 & 0.124 & 0.109 \\
V5  &19:53:54.00&+18:47:03.37 & 12.70 &09.51 & 8782.2161& 161.& ... & ... &...\\
P2  &19:53:57.10&+18:45:46.98 & 17.70 &16.64 & 8781.1682 &0.367188 & --& 0.416& 0.378 \\
P3  &19:53:50.84&+18:47:51.55 & 18.99 &17.63 &8406.2069 & 0.37386 & 0.370362& 0.156& 0.071 \\
P4  &19:53:48.98&+18:47:49.87 & 17.69 &16.85 & 8407.1857 &0.556154 & 0.555352 & 0.485& 0.484 \\
P5  &19:53:34.28&$+$18:44:05.00 & 17.72 &16.77 & 8407.1975 & 0.404380 & 0.405700 & 0.162 & 0.189 \\
P6  &19:53:47.97&$+$18:47:55.84 & 16.31 &15.69 & 8406.2079& 0.0500 & 0.053767 & 0.053 & 0.047 \\
P7  &19:53:48.37&$+$18:47:47.20 & 19.06 &18.36 &8782.1583 & 0.0582 & 0.079270& ... & ...\\
P9  &19:53:48.61&$+$18:45:45.15	& 17.49 &16.74 & 8782.1170& ...&0.072603&...&...\\
P10 &19:54:01.73&$+$18:47:18.32 & 15.35 &14.59 & 8781.1247 &0.76842 & 0.633149& 0.366& 0.363 \\
P15 &19:53:53.14&$+$18:52:23.65 & 18.28 &17.20 & 8781.1734 &...    & 0.358396 & 0.206 & 0.186  \\
P16 &19:53:58.96&$+$18:49:28.52 & 16.63 &15.90 & 8781.1541& 0.3970 & 0.371536& 0.030 & 0.039  \\
P17 &19:54:00.25&$+$18:52:35.25 & 19.39 &18.34 & ...&0.2745 & ...&...&...\\
P21 &19:53:25.55&$+$18:51:17.39 & 16.14 &15.07 & 8406.1894& 0.353 & 0.359764 & 0.392& 0.403  \\
P22 &19:53:48.67&$+$18:45:57.93	& 15.91 &15.20 & 8407.1101&0.0500 &0.115286&...&...\\

\hline
\end{tabular}
\end{table*}
\section{{Census of variable stars in M71} }
\label{sec:var_stars}


There is a large number of variable stars in the wider field of M71. However, it can be argued on the basis of recent membership approaches that many of them do not pertain to the cluster. The census of variable stars in M71 has been notoriously confusing, mainly due to the different nomenclatures used for the same variables in the reports of new discoveries of variables in the field (e.g. \citet{Arp1971}; \citet{Hodder1992}; \citet{YanMateo1994A};\citet{ParkNem2000}; \citet{Rucinski2000}; \citet{McCormac2014}). The several lists of variables and naming conventions have been clearly sorted out in the notes for M71 of CVSGC, March 2012 edition \citep{Clement2001}. In the end, \cite{Clement2001} lists the variables V1-V6 and P1-P23 as unequivocal in the \citet{ParkNem2000} numbering system. In the present paper, we will use this list of variables as the starting point towards the updating of the cluster member variables; we shall make an effort to measure their light curves and argue on the possible cluster membership status of each variable. 

\begin{figure}
  \centering
  \includegraphics[width=8.0cm]{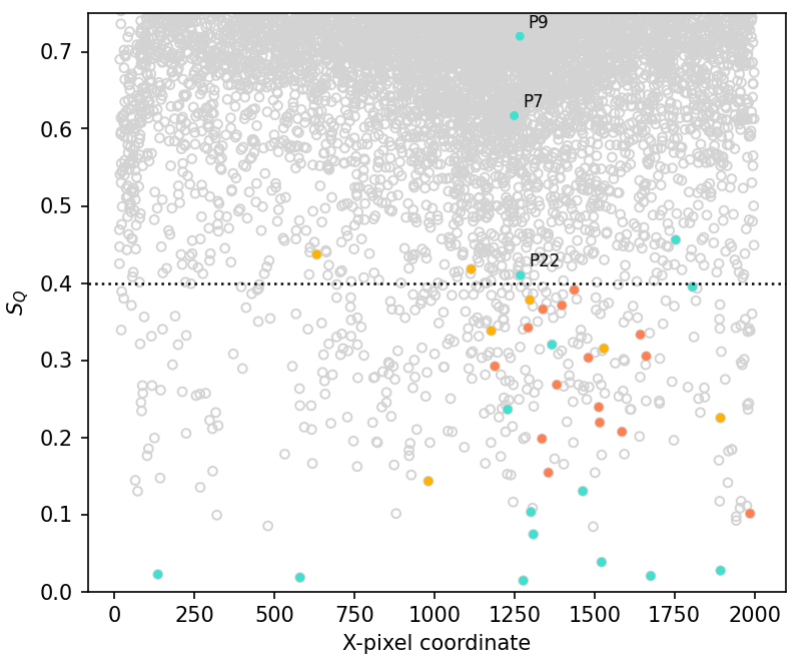}
  \caption{Minimum string–length statistic $S_Q$ for stars with $V$–band time series. The dashed line marks the empirical detection threshold at $S_Q=0.4$. Trial periods were scanned within $P_{\min}=0.01$~d, $P_{\max}=0.80$~d, and $\Delta P=10^{-6}$~d. All point sources measured in this work are shown as gray open circles; previously known variables from the CVSGC are plotted in turquoise, new detections from the string–length method in coral, and those identified through the median-based screening in yellow.}
  \label{fig:SQ}
\end{figure}

Due to the resolution of our images and the sky conditions during the observations, we were unable to resolve the light curves of some variables in the central regions of the cluster or in very tight star groups. In Table \ref{tab:clement_comparison} we list the stars that we could measure and their light curve properties. Their light curves are displayed in Fig. \ref{fig:Clement_Variable}. 
Given their time baseline, cadence, and/or signal-to-noise, the periods and amplitudes of three variable stars in the CVSGC (V2, V5, and P17) could not be estimated and/or refined with the data used in the present work. 
No evident variation is seen in the light curve of P9, and a very mild suggestion of variability is detected in $V$ for P7 and P22.
These stars are faint and are located in the very central region of the cluster.
In the $S_{Q}$ string-length statistic diagram of Fig.\ref{fig:SQ} (see Sect.~\ref{sec:string_length}), these three stars have large $S_{Q}$ indices, i.e., a long string-length value, meaning a very unlikely variability.
We conclude that 
with our present photometry we are unable to confirm the variability of P9. We recover low-amplitude variation in P7 at a level comparable to that reported by \citet{Hodder1992}.
We refer the reader to the individual notes on these three stars contained in Appendix A. Nevertheless, not all these variables are cluster members.
For each star, we have indicated the membership status assigned following two independent approaches: (a) from the method of \citet{Bustos2019} described in Sect.~\ref{sec:membership} and (b) from the membership code listed in the CVSGC from the analysis of \citet{PrudilArellanoFerro2024}.
For those cluster member stars presently listed with the prefix `P', we suggest numbering them using a prefix `V', following the numbering system used in the CVSGC.

In the images from Hanle Observatory, we were able to measure 7327 point sources in the FoV. Then, we performed a systematic search for variability to discover new variables or compare them with previous detections in the literature. We proceeded as described below.

We cross–matched our sample with the CVSGC  (Sect.~\ref{sec:clement}), and refined the periods where warranted. In addition, we conducted a two–pronged search for variability on all stellar sources detected within our field of view: (i) a periodogram-free phase–ordering approach based on the \emph{string–length} statistic $S_{Q}$ (Sect.~\ref{sec:string_length}), and (ii) a robust intersite consistency screen, comparing Hanle and SPM time series (Sect.~\ref{sec:robust_screen}). 



\subsection{Cross-match with previously reported variables}
\label{sec:clement}

We cross-matched our sample with the CVSGC. For M71, the catalogue lists 29 entries spanning SX Phe, eclipsing systems, and long-period variables. Using our time-series photometry, we refined the periods for ten of these previously known variables and redefined their classifications. The adopted values are reported in Table~\ref{tab:clement_comparison}, and the corresponding phase-folded $VI$ light curves are shown in Fig.~\ref{fig:Clement_Variable}, where dark teal points represent the Hanle data and black points correspond to the SPM observations.

Final periods were refined using the Phase Dispersion Minimization (PDM; \citealt{Stellingwerf1978}) algorithm, implemented through the \texttt{PyAstronomy} package\footnote{\url{https://github.com/sczesla/PyAstronomy}}
. This method identifies the true period as the minimum of the phase–dispersion statistic within the explored frequency range. Each light curve was then phased with the adopted period to verify the coherence of the modulation in both filters. Period uncertainties were estimated using the Generalised Lomb–Scargle (GLS; \citealt{Zechmeister2009}) periodogram, also computed via \texttt{PyAstronomy}, by scanning a narrow window around the PDM solution and adopting the period error provided by the GLS when both methods converged to the same value. This approach follows the procedure described by \citet{Cortes2023}. 


%


\begin{figure*}
  \centering
  \includegraphics[width=1.03\hsize]{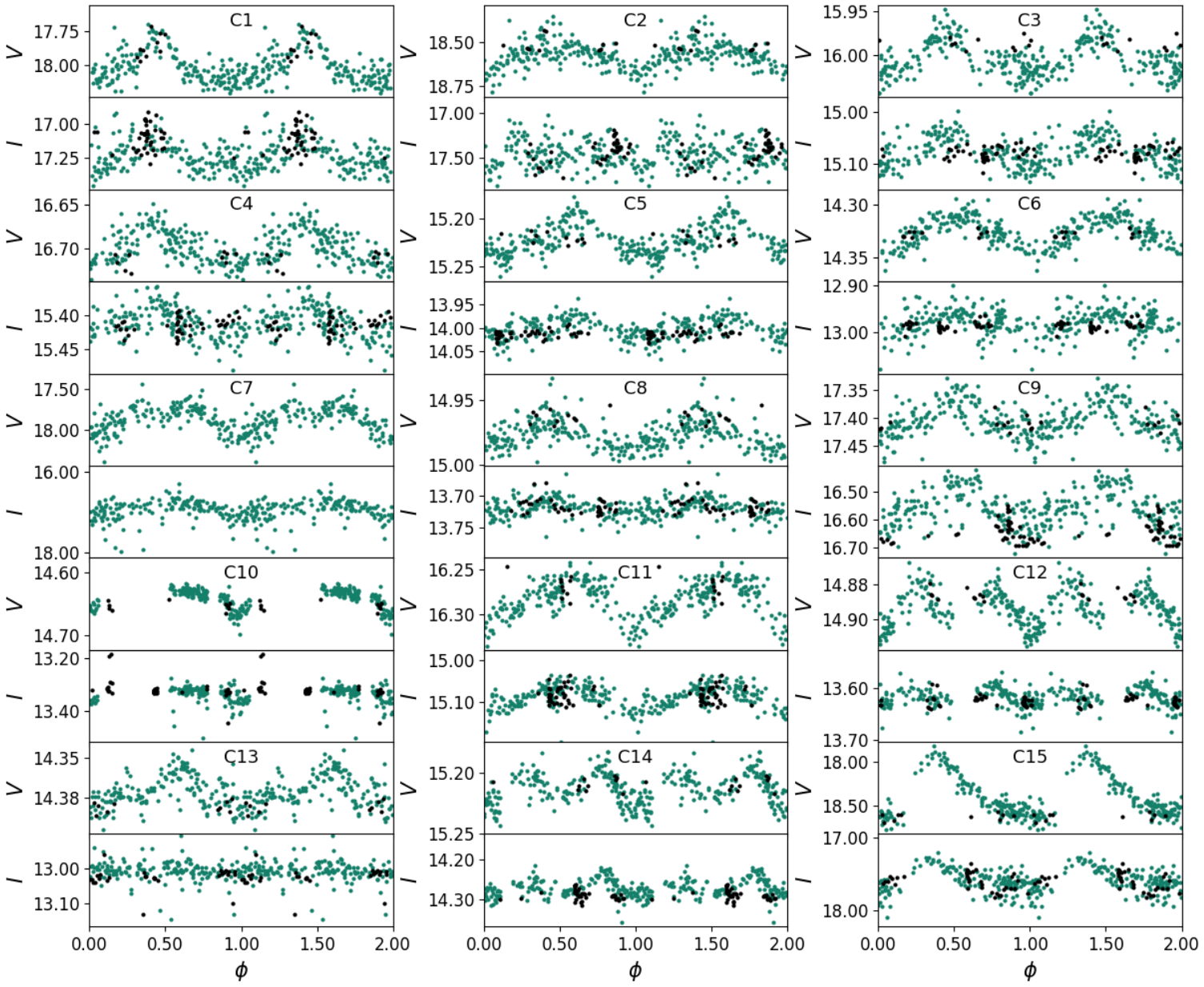}
  \caption{Phase–folded $V$ (top) and $I$ (bottom) light curves for the variables identified with the string–length method. Adopted periods and amplitudes are those reported in Table \ref{tab:all_variable}}. The colour code is as in Fig. \ref{fig:Clement_Variable}.
  \label{fig:SL_lightcurves}
\end{figure*}

\subsection{String–length Method}
\label{sec:string_length}

We identified 15 candidate variable stars using the string–length method \citep{Burke1970,Dworetsky1983}, applied independently to the $V$– and $I$–band time series. For each source, the light curve was phase–folded over a dense grid of trial periods defined within $P_{\min}=0.01$~d and $P_{\max}=0.80$~d, with a step size of $\Delta P=1\times10^{-6}$~d. At each trial period, we computed the string-length statistic $S_Q$ by ordering the measurements by phase and summing the normalised lengths of the consecutive segments joining adjacent points, including the closure across phase~1. The true period corresponds to the minimum $S_Q$ value, reflecting the highest phase coherence in the folded light curve. This approach is particularly effective for detecting non–sinusoidal variations, such as those exhibited by SX~Phe, $\delta$~Sct, and EW/EA systems, where Fourier-based periodograms can be less sensitive. We adopted an empirical detection threshold of $S_Q=0.4$, below which candidates were visually inspected to confirm phase coherence and consistency across both filters. The distribution of minimum $S_Q$ values and the adopted threshold are illustrated in Fig.~\ref{fig:SQ}, where all sources detected in this work are shown as gray open circles. Detected previously known variables from the CVSGC are marked in turquoise, the 15 variables identified with the string–length method are shown in coral, and 6 stars detected through the median-based screening method (see Sect.~\ref{sec:robust_screen}) are displayed in yellow.

The general information of the 15 detected candidate variable stars, named C1-C15, is summarised in Table~\ref{tab:all_variable} and their light curves are displayed in Fig.~\ref{fig:SL_lightcurves}. The columns are as follows: \emph{(1)} internal identification number; \emph{(2-3)} right ascension and declination (J2000.0) from \textit{Gaia}~DR3; \emph{(4-5)} mean standard magnitudes $\overline{V}$ and $\overline{I}$; \emph{(6)} ``Epoch min.'' denotes the epoch of minimum light adopted for the phase-folded light curves; \emph{(7)} refined period; and \emph{(8-9)} peak-to-peak amplitudes in the $V$ and $I$ bands, respectively. Phase-folded light curves for the 15 string-length detections are shown in Fig.~\ref{fig:SL_lightcurves}.



\begin{figure*}
  \centering
  \includegraphics[width=\hsize]{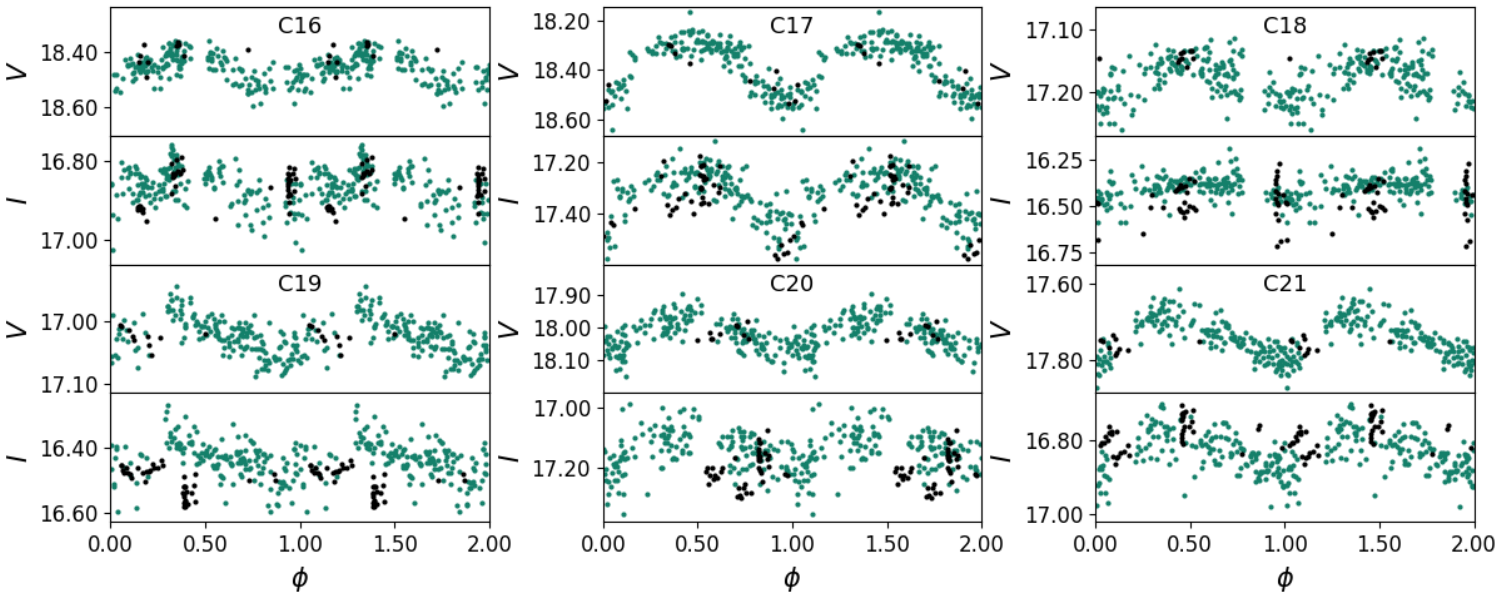}
  \caption{Phase–folded $V$ (top) and $I$ (bottom) light curves for the 6 variables identified with the median-based inter-site screen. Periods are the adopted PDM solutions (seeded by string–length when available); amplitudes are derived from GLS as detailed in Sect.~\ref{sec:robust_screen} and listed in Table \ref{tab:all_variable}.} The colour code is as in Fig. \ref{fig:Clement_Variable}.
  \label{fig:RS_lightcurves}
\end{figure*}

\begin{table*}
\centering
\caption{Variables detected in the field of M71 (see Sections~\ref{sec:string_length} and \ref{sec:robust_screen}).}
\label{tab:all_variable}
\begin{tabular}{lcccccccc}
\hline
ID  & RA  & Dec   &  $\overline{V}$ &  $\overline{I}$ & Epoch min.& $P_{V}$ & $A_V$  & $A_I$ \\
 &(J2000)&(J2000)& (mag)  & (mag)  &2450000.+&(day)  &  (mag)  & (mag)  \\
\hline
C1 &19:53:46.92&$+$18:44:35.36& 18.03 &17.26&8406.1351 &0.163327 &0.229&0.155\\
C2 &19:53:49.08&$+$18:44:33.69& 18.56 &17.41&8406.1623 &0.190200 &0.141&0.242\\
C3 &19:53:50.26&$+$18:48:54.27& 16.02 &15.09&8406.0623 &0.201756 &0.042 &0.060\\
C4 &19:53:50.31&$+$18:48:41.64& 16.68 &15.39&8406.1963 &0.231669 &0.038 &0.038 \\
C5 &19:53:50.67&$+$18:48:47.73& 15.21 &13.99&8406.1593 &0.199206 &0.034 &0.032 \\
C6 &19:53:51.20&$+$18:48:30.96& 14.30 &12.95& 8406.1640 &0.346422&0.032 &0.036 \\
C7 &19:53:51.48&$+$18:46:00.57& 17.85 &16.87&8405.9450 &0.285495 &0.315&0.332\\
C8 &19:53:52.15&$+$18:45:47.92& 14.96 &13.70&8406.1496 &0.164935&0.020 &0.015 \\
C9&19:53:53.15&$+$18:47:10.38& 17.42 &16.57&8406.1620 &0.240929 &0.0579 &0.153\\
C10&19:53:53.63&$+$18:44:10.10& 14.62 &13.31&8406.3557 &0.911779&0.036 &0.066 \\
C11&19:53:53.83&$+$18:46:14.28& 16.27 &15.08&8406.2264 &0.367880&0.042 &0.049 \\
C12&19:53:55.35&$+$18:47:24.74& 14.87 &13.60 &8405.9796 &0.381886&0.024 &0.026 \\
C13&19:53:56.53&$+$18:47:11.70& 14.34 &12.98&8406.0905 &0.263071&0.018 &0.031 \\
C14&19:53:56.91&$+$18:47:20.76& 15.23 &14.28&8405.9105 &0.406560 &0.022 &0.035 \\
C15&19:54:03.81&$+$18:49:53.79& 18.31 &17.57&8406.1405&0.199353&0.557&0.328\\
C16&19:53:35.42&$+$18:45:00.70& 18.40 &16.81&8406.2552 &0.566049&0.106 &0.075 \\
C17&19:53:42.70&$+$18:45:58.15& 18.36 &17.31&8406.1796&0.19056 &0.242 &0.196\\
C18&19:53:45.44&$+$18:45:28.57& 17.21 &16.43&8406.2290 &0.326517&0.059 &0.088\\
C19&19:53:46.84&$+$18:46:59.97& 17.07 &16.48&8406.1798 &0.321780&0.057 & 0.090\\
C20&19:53:49.37&$+$18:48:00.38& 18.04 &17.16&8405.9377 &0.374476&0.105 & 0.104\\
C21&19:54:01.45&$+$18:43:17.38& 17.76 &16.85&8406.1096 &0.324253&0.111 & 0.099 \\

\hline
\end{tabular}
\end{table*}

\subsection{Median-based inter-site variability screen method}
\label{sec:robust_screen}

To complement the string--length search, we performed a \emph{median}-based site--to--site consistency test designed to expose low-amplitude variables by comparing robust location estimates between the Hanle and SPM campaigns. Because the Hanle data comprise two well-separated runs, each Hanle light curve was split into two temporal groups (G1, G2) using a one-dimensional $k$--means clustering on the observation times. For each group, and for SPM, we computed the \emph{median} magnitude $\tilde{m}$ in $V$ and $I$ after iterative clipping based on the median absolute deviation (MAD), using the standard scaling $\mathrm{MAD}\times1.4826$ to approximate Gaussian $\sigma$ \citep[e.g.,][]{Hoaglin1983,Rousseeuw1993}. 

From these medians we formed (i) the absolute inter-site offsets per group,
$|\Delta V_{\mathrm{G1}}|=|\tilde{V}_{\mathrm{SPM}}-\tilde{V}_{\mathrm{Hanle,G1}}|$, 
$|\Delta V_{\mathrm{G2}}|$, 
$|\Delta I_{\mathrm{G1}}|$, 
and $|\Delta I_{\mathrm{G2}}|$;
(ii) the internal Hanle consistency metrics 
$|\Delta V_{\mathrm{Hanle}}|=|\tilde{V}_{\mathrm{Hanle,G2}}-\tilde{V}_{\mathrm{Hanle,G1}}|$ 
and $|\Delta I_{\mathrm{Hanle}}|$; 
and (iii) the effective sample sizes after clipping. 
A source was provisionally flagged as a \emph{robust-screen candidate} when at least one SPM--Hanle median difference exceeded an empirical threshold of 0.05\,mag, and---optionally---when the internal Hanle median difference exceeded an independent control threshold.

To ensure the reliability of the detected periodic signals, we followed the combined GLS+PDM procedure described by \citet{Cortes2023}. The period of each candidate was first determined using the PDM and subsequently validated with the GLS. Both techniques yielded consistent results within their formal uncertainties. Period and amplitude errors were estimated from the GLS fits, following the same procedure adopted for the CVSGC variables in Sect.~\ref{sec:clement}. As an additional statistical filter, only candidates displaying a significant GLS peak above the 0.1\% false–alarm probability (FAP) level in \emph{both} $V$ and $I$ bands were retained. The FAP quantifies the probability that a detected periodicity arises purely from random noise rather than from a genuine signal; thus, the adopted 0.1\% threshold corresponds to a confidence level greater than 99.9\%, ensuring that all accepted detections are statistically robust \citep{ZechmeisterKurster2009}.

The final stage of verification consisted of a visual inspection of the phase–folded light curves. Each candidate was examined to confirm that its morphology was consistent with genuine stellar variability, such as coherent periodic modulation, smooth amplitude changes, and reproducibility across both filters. Through this multistage filtering process, we identified six candidate variable stars that meet all criteria. We have named these stars C16-C21. Their general properties—coordinates, photometry, periods, amplitudes, and membership flags—are listed in Table~\ref{tab:all_variable}, while their phase–folded $V$ and $I$ light curves are shown in Fig.~\ref{fig:RS_lightcurves}, where the dark teal points correspond to the Hanle data and the black points to the SPM observations.

We note that the median-based inter-site screening method did recover several variable stars previously reported in the CVSGC, as well as some objects already identified through the string-length analysis. However, since these variables were independently detected by other approaches within this work and to avoid duplication in the variable-star census, we report here only those sources that constitute independent detections of the median-based method. In addition, we retain exclusively the candidates that satisfy the adopted statistical significance requirement, namely a generalized Lomb--Scargle false-alarm probability below the imposed threshold. Consequently, the six stars presented in this section represent new variables not previously reported by the preceding methods and define the specific contribution of this procedure to the overall census of variable stars in the field of M71.




\begin{figure}
\begin{center}
\includegraphics[width=9cm]{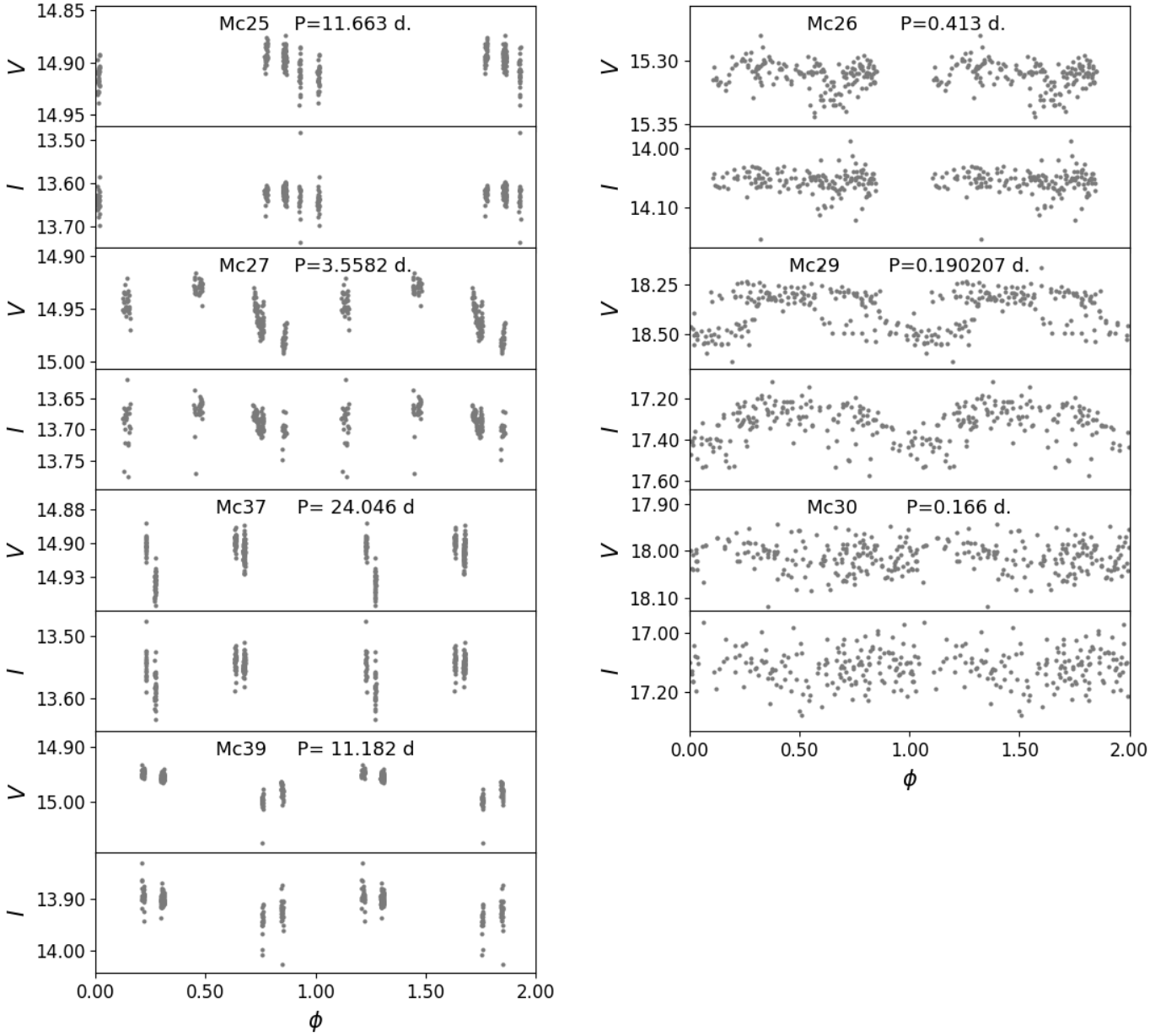}
\caption{Phased light curves folded with the periods reported by \citet{McCormac2014}. The variables shown in this figure include stars identified by those authors as likely cluster members, together with C17, which is classified here as a non-member according to the membership analysis described in Sect.~\ref{sec:membership}. Mc27 and Mc29 from \citet{McCormac2014} correspond to our sources C7 and C17, respectively. }
\label{fig:mc_members}
\end{center}
\end{figure}

\begin{figure}
\begin{center}
\includegraphics[width=6cm]{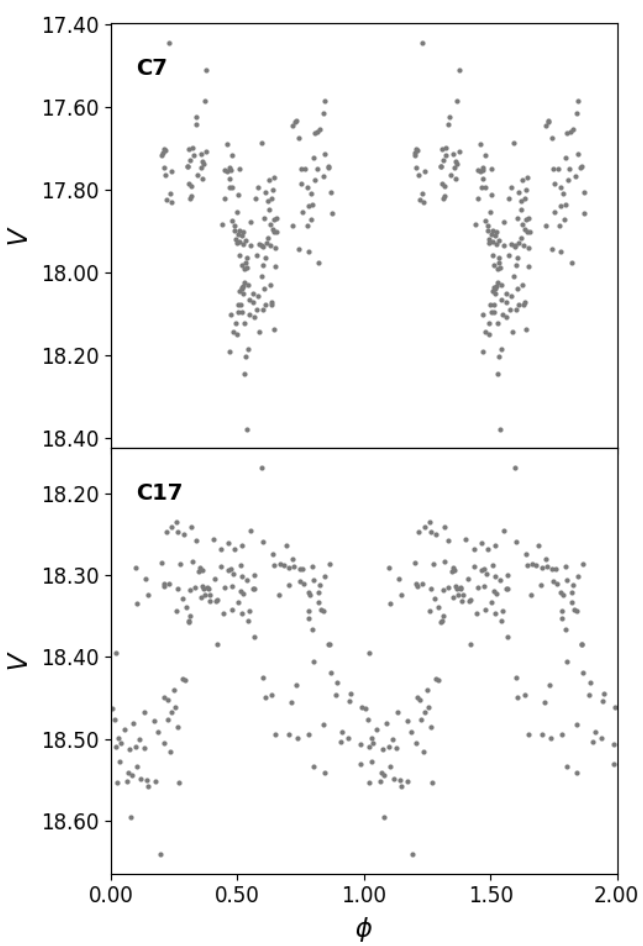}
\caption{Phased light curves of C7 (V0467 Sge) and C17 (V0456 Sge) folded with the periods reported by \citet{Watson2006}. In that work, V0467 Sge was classified as a $\gamma$ Dor star with a period of 0.779270~d, whereas V0456 Sge was classified as an SX Phe star with a period of 0.190207~d.}
\label{fig:W_S}
\end{center}
\end{figure}

\begin{figure}
\begin{center}
\includegraphics[width=8cm]{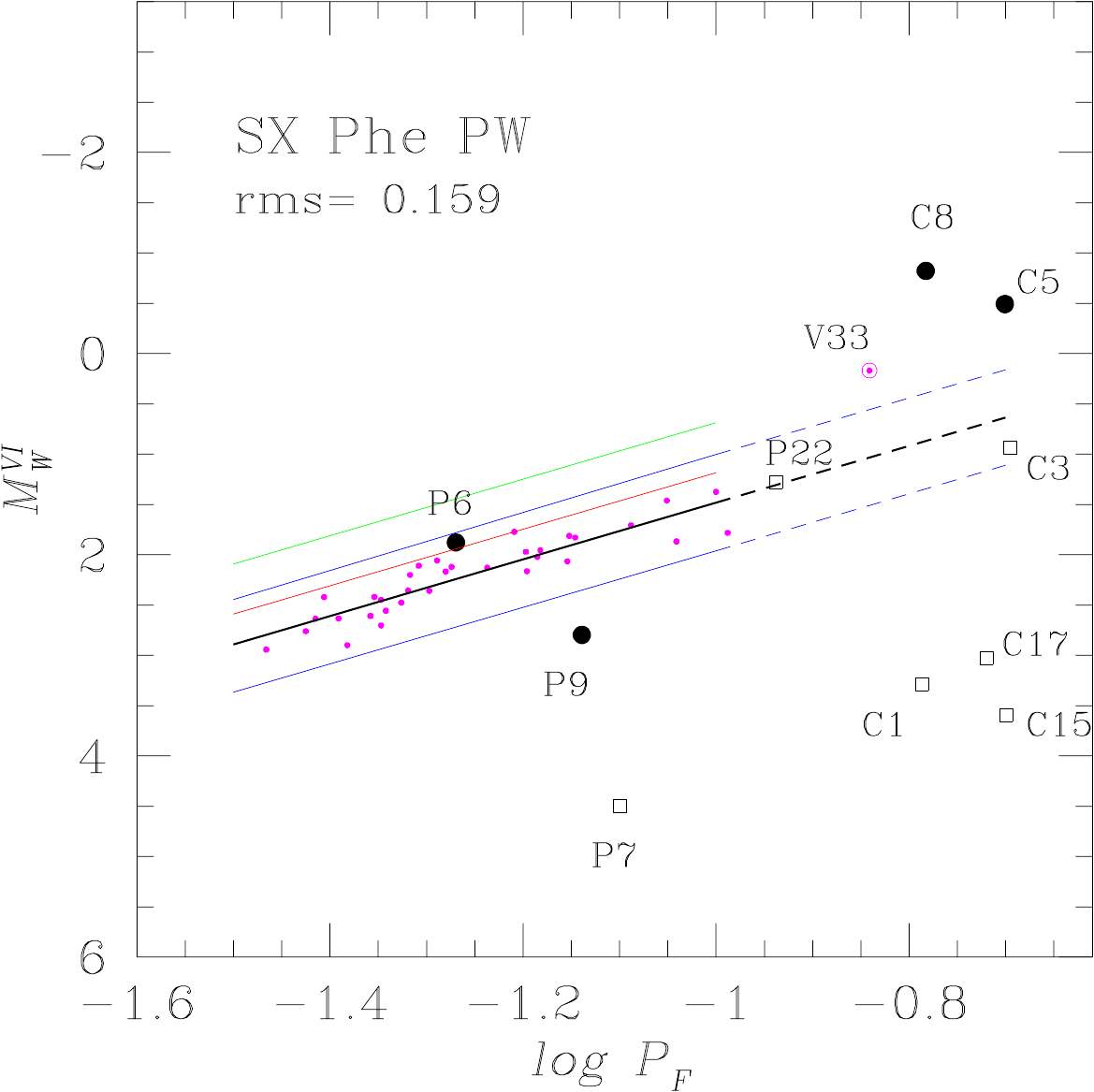}
\caption{Extinction-free PW relation for SX Phe stars in the $VI$-bands. Small magenta points represent a sample of 34 SX Phe from Table 1 of \citet{Ngeow2023}. Circled point is V33 in NGC 6341, not included in the fit and noted as an outlier by \citet{Ngeow2023}. The black line fit and extrapolation to $log~P= -0.8$ ($P= 2 d$), is represented by $M^{VI}_W = -2.815(\pm 0.226) log~P - 1.334 (\pm 0.294)$; rms = 0.159. Blue lines are the $3\sigma$ boundaries of the dispersion. This suggests P6, P22 and probably P9 as SX Phe stars. P7 is not an SX Phe. See Sect.~\ref{VarTypes} for a discussion.}
\label{fig:SX_PL}
\end{center}
\end{figure}

\begin{figure}
\begin{center}
\includegraphics[width=8cm]{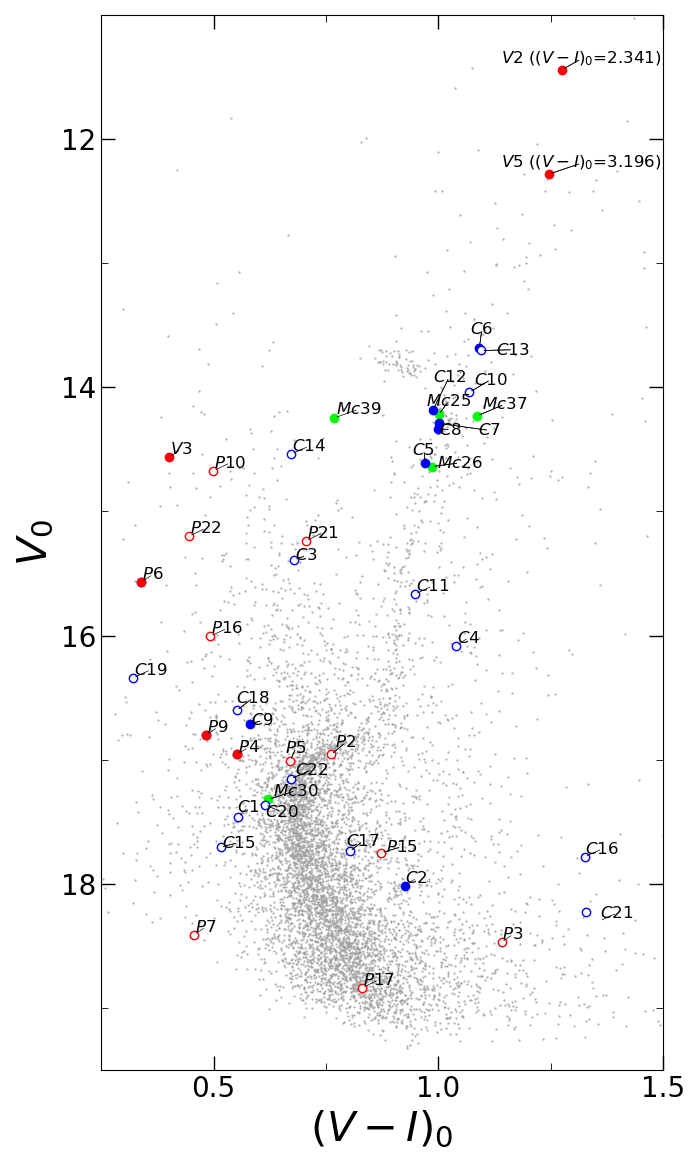}
\caption{Dereddened CMD of M71 showing the positions of previously known and newly identified variable stars. Filled circles represent stars classified as cluster members, while open circles denote likely field stars. Red circles mark the variables listed in the Clement catalogue, and green circles indicate those discovered by \citet{McCormac2014} that we identified as cluster members and have labeled with the prefix `Mc'. Blue circles show the new candidate variables identified in this study. Note that our sources C7 and C17 correspond to Mc27 and Mc29 from \citet{McCormac2014}, respectively. Also note that, given the scale of the plot and the marker sizes, some stars may appear to overlap. For clarity in the visualization, the positions of stars V2 and V5 have been artificially shifted towards the blue. Their original dereddened colors are indicated in the labels. See Sect.~\ref{sec:var_stars} for details.}
\label{fig:cmd_var}
\end{center}
\end{figure}

\begin{figure*}
  \centering
  \includegraphics[width=\hsize]{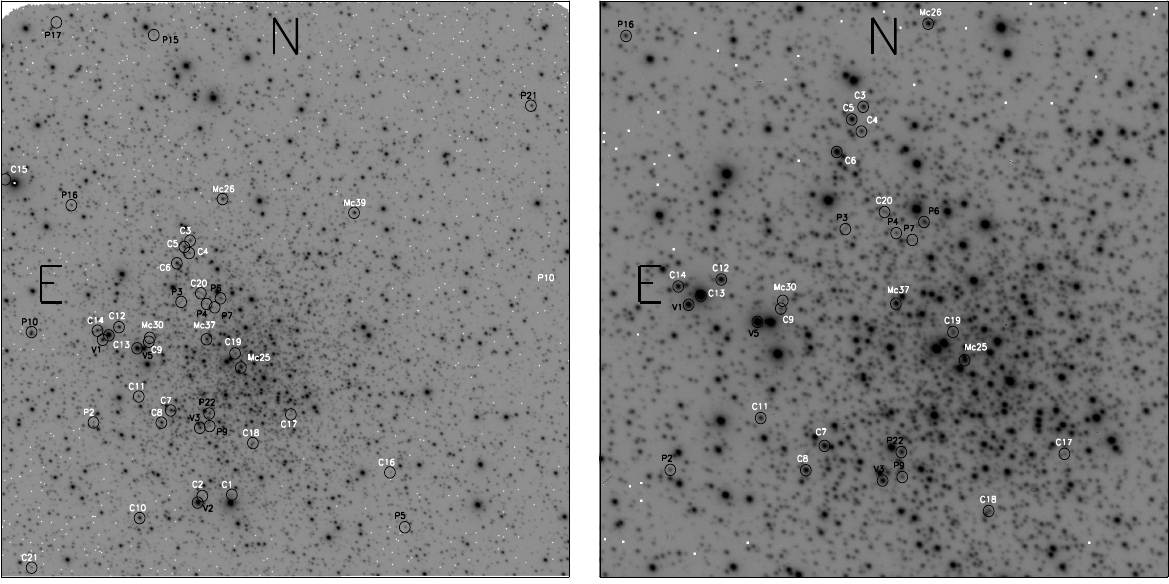}
  \caption{Identification chart of the variable stars in the field of M71. The left panel is approximately 10.0$\arcmin$ $\times$ 10.0$\arcmin$. The panel to the right is an expansion of the central region of the cluster and it is about 5.0$\arcmin$ $\times$ 5.0$\arcmin$. Stars labeled in black are known variables listed in the CVSGC. Those labeled in white are candidate variables reported in this work (C1-C21) and six variables from \citet{McCormac2014} that we found to be likely cluster members and have been identified with the prefix `Mc'. See Sect.~\ref{sec:var_stars} for a discussion.}
  \label{fig:ID}
\end{figure*}

\begin{table*}
\centering
\caption{Census of Variable stars in the field of M71}
\label{tab:CENSUS}
\begin{tabular}{lcccc c lcccc}
\hline
$\rm ID$ & New & memb & memb & Type & & ID & New & memb & memb & Type \\
         & name & B\&C & P\&AF &      & &    & name & B\&C & P\&AF &      \\
\hline
V2&   &m&m & SR& \rule{0.25pt}{2.2ex} &C1& &f & --& SX?\\
V3&   &m&m & EB& \rule{0.25pt}{2.2ex} &C2&V12 &m& --& Bin:\\
V5&   &m&m & SR& \rule{0.25pt}{2.2ex} &C3$^*$&V13&f/m & --& SX Phe\\
P2& V7&f&m & EW& \rule{0.25pt}{2.2ex} &C4& &f/m & --& RRc?\\
P3& V8&f&m & EW& \rule{0.25pt}{2.2ex} &C5&V14 &m & --& SX Phe\\
P4& V9   &m&m & EA& \rule{0.25pt}{2.2ex} &C6&V15 &m & --& RRc?\\
P5&   &f&f & EW& \rule{0.25pt}{2.2ex} &C7 (Mc27$^{**}$)&V16 &m & --& EB\\
P6& V10  &m&m & SX Phe& \rule{0.25pt}{2.2ex} &C8&V17 &m & --& SX Phe\\
P7&   &f&u & SX?& \rule{0.25pt}{2.2ex} &C9&V18 &m & --& ?\\
P9& V11  &m&m & SX?& \rule{0.25pt}{2.2ex} &C10& &f & --& EA\\
P10&    &f&f & EW & \rule{0.25pt}{2.2ex} &C11& &f & --& ?\\
P15&   &f&f & EB?& \rule{0.25pt}{2.2ex} &C12&V19 &m & --& EW\\
P16$^*$&   &f&f & RRc?& \rule{0.25pt}{2.2ex} &C13& &f & --& ?\\
P17&    &f&f & EB & \rule{0.25pt}{2.2ex} &C14& &f & --& EW?\\
P21&   &f&f & EW& \rule{0.25pt}{2.2ex} &C15& &f & --& SX?\\
P22$^*$&   &f&f & SX Phe& \rule{0.25pt}{2.2ex} &C16& &f & --& ?\\
-- & --&--&--&-- & \rule{0.25pt}{2.2ex} &C17 (Mc29$^{**}$)& &f & --& SX?\\
-- & --&--&--&-- & \rule{0.25pt}{2.2ex} &C18& &f & --& RRc?\\
-- & --&--&--&-- & \rule{0.25pt}{2.2ex} &C19& &f & --& RRc?\\
-- & --&--&--&-- & \rule{0.25pt}{2.2ex} &C20& &f & --& RRc?\\
-- & --&--&--&-- & \rule{0.25pt}{2.2ex} &C21& &f & --& RRc?\\  
-- & --&--&--&-- & \rule{0.25pt}{2.2ex} &Mc25$^{**}$& &m & --& \\
-- & --&--&--&-- & \rule{0.25pt}{2.2ex} &Mc26$^{**}$& &m & --&var?\\
-- & --&--&--&-- & \rule{0.25pt}{2.2ex} &Mc30$^{**}$& &m & --& var?\\
-- & --&--&--&-- & \rule{0.25pt}{2.2ex} &Mc37$^{**}$& &m & --& \\
-- & --&--&--&-- & \rule{0.25pt}{2.2ex} &Mc39$^{**}$& &m & --& \\
\hline
\end{tabular}
           \center{* New possible classification. C3 and P22 are likely cluster members according to the PW analysis shown in Fig. \ref{fig:SX_PL}. Likewise, C5 and C8 are probably not cluster members after their position in the PW diagram. }
    \center{** Variable stars according to \citet{McCormac2014}, where these objects are listed as v25, v26, v27, v29, v30, v37 and v39. In particular, C7 and C17 correspond to Mc27 and Mc29, respectively. We did not confirm the variability of v26 (Mc26) and V30 (Mc30)}. See Appendix A for further comments. 
                      \center{: Uncertain classification}
\end{table*}

\subsection{Variable types}
\label{VarTypes}

For each variable listed in Tables \ref{tab:clement_comparison} and \ref{tab:all_variable}, its position in the CMD, period, and light curve morphology may, in most cases, suggest its variable type. In Table \ref{tab:CENSUS} we summarise the membership status and, when possible, the likely variable type of all variables detected in the field of M71. On the left side of the table, we list the known variables and their types as given in the CVSGC. On the right side, we include the newly identified variables (C1--C21) and five Mc stars that were announced as variables by \citet{McCormac2014} and that, apparently, are cluster members. A cross-check with \textit{Gaia}~DR3 shows that none of the 21 variable candidates identified in this work is flagged as a variable source there.\\
\\
C7 and C17 correspond to the variables V0467 Sge and V0456 Sge previously reported by \cite{Watson2006}, and to Mc27 and Mc29 in \cite{McCormac2014}, respectively. These stars are therefore considered previously known variables. Their phased light curves, together with the periods reported by \cite{McCormac2014} and \cite{Watson2006}, are presented in Figs. \ref{fig:mc_members} and ~\ref{fig:W_S}, respectively. In both cases, our adopted periods provide a more consistent phased representation of the variability in our data, and are therefore preferred as refined period estimates. However, we did not detect clear variability for stars Mc26 and Mc30 in our photometry.\\
\\
Some of the stars marked with “?” in Table~\ref{tab:CENSUS} show periods, amplitudes, and light-curve morphologies that do not allow a secure classification into any classical variability type. Similar cases have been reported in recent globular-cluster studies \citep{Cortes2023, Cortes2026}. We therefore regard these objects as stars whose variability type remains unresolved.\\
\\
The variable types assigned in Table \ref{tab:CENSUS} for these stars are discussed in this section. Although the position of a star in the CMD, its period, and its light-curve morphology may in most cases suggest its variable type, special attention is given to the RRc and SX candidates through the log P–amplitude and log P–Wesenheit relations, respectively, as described below.\\
\\
Due to their period and light curve morphology, seven stars are considered possible RRc stars (P16, C4, C6, C18, C19, C20, and C21). Their position in the log P-Amplitude diagrams, both in the $V$ and $I$-bands, follows the expected distribution of RRc stars of the Oo I type (figure not shown). None of these stars are cluster members but are instead background stars, as can also be noted from their position in the CMD, well below the HB region.\\
\\
For the SX Phe candidates, i.e., stars with $P$ < 0.2 d, we approached the discussion of their nature from their positions in the log P vs. Wesenheit index plane. The reddening-free Wesenheit index is defined as: $$W^{ri} =  r - R_{ri} (r-i)$$  with $R_{ri} = 4.051$ \citep{Madore1991}. In the $VI$-bands, the index takes the form:
$$W^{VI} = V - R_{VI} (V-I)$$ with $R_{VI} = 2.217$ (\citealp{Ngeow2022}; see their Table 4.)

We therefore used the $gri$ photometry of SX Phe stars in globular clusters from Table 1 of \citet{Ngeow2023} and converted it into $V$ and $I$ magnitudes employing the transformation equations from \citet{Tonry2012}  (see their Table 6), for 34 stars with $i$-band photometry available, and calculated their  $W^{VI}$. Then, the absolute magnitude of this Wesenheit index was obtained via $M^{VI}_W = W^{VI}-5logD$+5. The distances of the host cluster were taken from \citet{Baumgardt2021}.

In the same way, we calculated the $M^{VI}_W$ reddening-free absolute magnitude for the candidate SX Phe stars by adopting the distance modulus 13.01 for M71 \citep{Baumgardt2021}.
In Fig. \ref{fig:SX_PL}, the small cyan points represent the SX Phe sample from  \citet{Ngeow2023}. The periods of all SX Phe pulsating in the first or second overtone were fundamentalised using the ratios $1O/P_F = 0.783$ and $2O/P_F = 0.571$ \citep{Arellano2011}. The linear fit $M^{VI}_W = -2.815~(\pm 0.226)~ log P - 1.334~(\pm 0.294)$; rms = 0.159 mag represents a new extinction-free PW relation for the $VI$-bands for SX Phe stars, not included in the work of \citet{Ngeow2023}.

This exercise confirms P6 as a true SX Phe star, consistent with the classification by \citet{ParkNem2000} and also identifies the new variables C5 and C8 as cluster member SX Phe stars.

At this point, we would like to recall that \citet{McCormac2014} reported 17 new variables, which they numbered v24-40. These stars were not included in the CVSGC because they were considered probably field stars. However, in the membership analysis of Sect.~\ref{sec:membership} stars v25, v26, v27, v30, v37 and v39 were found to be likely cluster members. To avoid confusion with the naming system employed in the CVSGC, we retained these stars as Mc25, Mc26, Mc27, Mc30, Mc37, and Mc39 in Table \ref{tab:CENSUS}. In Fig.~\ref{fig:mc_members}, we show the phase-folded light curves of these variables using the periods derived by \citet{McCormac2014}. We did not confirm the variability of v26 and v30 in our photometry. 

As before, for those stars that are likely members, we suggest numbering them according to the CVSGC convention, i.e., with a prefix `V'. These suggested names are given in columns 2 and 7 of Table \ref{tab:CENSUS}. In conclusion, Table \ref{tab:CENSUS} contains those variable stars in the field of M71, and flags those that very likely pertain to the cluster, and therefore represent the census of the variable stars in M71. In Fig.~\ref{fig:cmd_var} we display all variables on the cluster CMD, and their sky positions on the finding chart in Fig.~\ref{fig:ID}.

\section{Conclusions}
\label{sec:conclusions}

In this work, we present a photometric study of the globular cluster M71 (NGC~6838) based on time-series observations in the \(V\) and \(I\) bands, obtained during two independent observing campaigns carried out at the San Pedro Mártir Observatory and the Indian Astronomical Observatory. Cluster membership was established using \textit{Gaia} DR3 proper motions, by identifying the population associated with M71 in proper-motion space and validating the selection through its spatial distribution and its consistency with the evolutionary sequences in the colour--magnitude diagram. In addition, a differential reddening correction was applied to improve the definition of the cluster's evolutionary sequences. This combination of criteria allowed us to construct a cleaned colour--magnitude diagram, suitable for estimating the fundamental cluster parameters through theoretical isochrone fitting

Based on the cleaned colour--magnitude diagram, the fundamental parameters of M71 were estimated through theoretical isochrone fitting using a Bayesian approach. The derived age, metallicity, mean reddening, and distance modulus are consistent, within their uncertainties, with previous determinations based on photometric and spectroscopic studies. These results confirm that M71 is a relatively young and metal-rich globular cluster and further support its association with the Galactic disc population.

In the analysis of stellar variability, an updated census of variable stars in the field of M71 was constructed using a complementary methodological strategy. The search combined a periodogram-free approach based on the string-length statistic, refined through phase-dispersion minimization, with an independent screening grounded on the inter-site consistency of robust medians and subject to a strict statistical significance criterion. As a result, periods and classifications of previously reported variables were refined, and 21 new variable stars were identified, representing independent detections of the applied methods and significantly expanding the census of variables in the cluster field.

The classification of the variable stars was based on the morphology of their light curves, the derived periods, and their location in the colour--magnitude diagram, which allowed us to identify predominantly eclipsing binary systems and short-period pulsating stars. In the case of the SX~Phoenicis candidates, their position in the period-Wesenheit relation was analysed using the adopted cluster distance modulus and reddening. This analysis enabled us to discriminate between cluster members and field stars, to confirm the nature of some previously reported variables, and to establish new SX~Phoenicis stars as \textit{bona fide} members of M71.


\section*{Acknowledgments}

AAF is grateful to DGAPA-UNAM for financial support through project  IN103024. We thank the support staff at both the IAO and CREST facilities operated by the Indian Institute of Astrophysics, Bangalore, and the SPM Observatory operated by the Institute of Astronomy, UNAM.

This work has made use of data from the European Space Agency (ESA) mission
{\it Gaia} (\url{https://www.cosmos.esa.int/gaia}), processed by the {\it Gaia}
Data Processing and Analysis Consortium (DPAC,
\url{https://www.cosmos.esa.int/web/gaia/dpac/consortium}). Funding for the DPAC
has been provided by national institutions, in particular the institutions
participating in the {\it Gaia} Multilateral Agreement.      
\\

{\bf DATA AVAILABILITY}

The data employed in this work shall be available in electronic
form in the Centre de Donnés astronomiques de Strasbourg data
base (CDS), and can also be shared on request to the corresponding
author.


\bibliographystyle{mnras}
\bibliography{example} 




\appendix

\section{Discussion on individual stars}
\label{appendix}
{\bf V3 (QU Sge).} This EB star was reported with a period of 3.7907~d by \citet{SawyerHogg1973}. The star was further studied by \citet{Jeon2006} who found that one of the components is an SX Phe variable. Our observations cover only a small portion of the light curve as shown in Fig. \ref{fig:Clement_Variable} where we have sketched the orbital solution from \cite{Jeon2006}. 

{\bf P2 and P3.} These two eclipsing binaries of the EW type were found to be cluster members by the analyses
of \citet{Vasiliev2021} and \citet{PrudilArellanoFerro2024} but were labeled as field stars by our analysis described in Sect.~\ref{sec:membership}.

{\bf P7.} The variability of this faint star was announced by \citet{Hodder1992} based on a 6-hour time span light curve, well phased with a period of 0.0582 d. It was considered a field dwarf Cepheid. However, its location in the Blue Straggler region led \citet{ParkNem2000} to suggest that it is an SX Phe star. These authors were unable to provide new photometry. Our more numerous data set and considerably larger time span data suggest P = 0.07927~d resulting into an equivalent variation to the one found by \citet{Hodder1992} in spite of having more scatter. We found the star to be likely a field star. The star could not be identified by \citet{McCormac2014}. 

{\bf P9.} It was reported as variable, probably of the SX Phe type, due to its position in the Blue Straggler region, by \citet{ParkNem2000}. Its  period could not be determined. Only small fractions of a noisy light curve were attained. In our photometry we were also unable to confirm the variability and hence the nature of this faint star, that according to our analysis is, however, a cluster member. No light curve was measured by \citet{McCormac2014}.

{\bf P22.} Although the star was retained as variable in the CVSGC, due to its variability on a time span of 7 hours on a single night \citep{ParkNem2000}, its variability was doubted by \citet{ParkNem2000}. It is often referred to as a low amplitude SX Phe star. We detected a mild variation with a period of 0.115286 d. Its position on the extinction-free PW relation for SX Phe (see Fig. \ref{fig:SX_PL}) suggests the star to be an SX Phe member of the cluster. 

{\bf Mc25, Mc26, Mc27, Mc30, Mc37 and Mc39.} These stars were reported by \citet{McCormac2014} as variable stars with the names v25, v26, v27, v30, v37 and v39. They were found to be likely cluster members in our analysis of Sect.~\ref{sec:membership}, however we could not confirm the variability of v26 and v30. Our stars C7 and C17 correspond to Mc27 and Mc29 respectively. \citet{McCormac2014} reported periods of 3.5582~d and 0.190207~d,respectively, and classified these stars as $\gamma$ Dor and SX Phe, respectively.  Both stars were measured in our data set from Hanle where our 4 nights data are not sufficient to explore a period of $\sim 3$ days in the case of C7. We found, however, that 0.285495~d represents well the variation in our data. Further photometry would be necessary to settle the period and variable type of C7. For the case of C17 we confirm the period 0.190207~d and consider this star as a field SX Phe star. 
\label{sec:appendix}


\bsp	
\label{lastpage}
\end{document}